\documentclass[12pt]{iopart}

\usepackage{iopams}  
\usepackage{color}
\usepackage[dvipsnames]{xcolor} 
\usepackage{tikz}
\usetikzlibrary{decorations.pathmorphing,patterns} 
\usepackage{psfrag}
\usepackage{graphicx}
\usepackage{stackrel}
\usepackage{amsfonts}
\usepackage{bm}          
\usepackage{amssymb}
\usepackage{perpage}
\MakePerPage{footnote}

\newcommand{\be}{\begin{equation}}
\newcommand{\ee}{\end{equation}}
\newcommand{\ba}{\begin{eqnarray}}
\newcommand{\ea}{\end{eqnarray}}

\begin{document}

\title[Critical points of Potts and O($N$) models from eigenvalue identities]
{Critical points of Potts and O($N$) models from eigenvalue identities in periodic Temperley-Lieb algebras}

\author{Jesper Lykke Jacobsen$^{1,2}$}
\address{${}^1$LPTENS, \'Ecole Normale Sup\'erieure -- PSL Research University, 24 rue Lhomond, F-75231
Paris Cedex 05, France}
\address{${}^2$Sorbonne Universit\'es, UPMC Universit\'e Paris 6, CNRS UMR 8549, F-75005 Paris, France} 

\eads{\mailto{jesper.jacobsen@ens.fr}}

\begin{abstract}

In previous work with Scullard, we have defined a graph polynomial $P_B(q,T)$ that gives
access to the critical temperature $T_{\rm c}$ of the $q$-state Potts model defined
on a general two-dimensional lattice ${\cal L}$. It depends on a basis $B$, containing $n \times m$
unit cells of ${\cal L}$, and the relevant root $T_{\rm c}(n,m)$ of $P_B(q,T)$ was observed to converge quickly
to $T_{\rm c}$ in the limit $n,m \to \infty$. Moreover, in exactly solvable cases there is no
finite-size dependence at all.

In this paper we show how to reformulate this method as an eigenvalue problem within the
periodic Temperley-Lieb algebra.
This corresponds to taking $m \to \infty$ first, so that the bases $B$ are semi-infinite cylinders of circumference $n$.
The limit implies faster convergence in $n$, while maintaining the $n$-independence in exactly solvable cases.
In this setup, $T_{\rm c}(n)$ is determined by equating the
largest eigenvalues of two topologically distinct sectors of the transfer matrix. 
Crucially, these two sectors determine the same critical exponent in the continuum limit,
and the observed fast convergence is thus corroborated by results of conformal field theory.

We obtain similar results for the dense and dilute phases of the O($N$) loop model, using now
a transfer matrix within the dilute periodic Temperley-Lieb algebra.

Compared with our previous study, the eigenvalue formulation allows us to double 
the size $n$ for which $T_{\rm c}(n)$ can be obtained, using the same computational effort.
We study in details three significant cases:
(i) bond percolation on the kagome lattice, up to $n_{\rm max} = 14$;
(ii) site percolation on the square lattice, to $n_{\rm max} = 21$;
and (iii) self-avoiding polygons on the square lattice, to $n_{\rm max} = 19$.
Convergence properties of $T_{\rm c}(n)$ and extrapolation schemes are studied in
details for the first two cases. This leads to rather accurate values for the percolation thresholds:
$p_{\rm c} = 0.524\,404\,999\,167\,439 (4)$ for bond percolation on the kagome lattice, and
$p_{\rm c} = 0.592\,746\,050\,792\,10 (2)$ for site percolation on the square lattice.

\end{abstract}

\noindent

\section{Introduction}

The question what makes a two-dimensional lattice model amenable to exact solution has attracted
considerable attention within the field of statistical mechanics. Most, but not all, solutions have been
found by the technique of integrability, in which the commutativity of an infinite family of transfer matrices
is ensured by requiring the Boltzmann weights to solve a set of cubic functional relations,
known as the Yang-Baxter equations \cite{Baxter_book}. 

Recent work has focussed on a construction called discrete holomorphicity (DH),
in which suitable correlation functions are required to satisfy a discrete version of the Cauchy-Riemann
equations \cite{Cardy_DH}. This leads to linear relations among the Boltzmann weights,
that express the conservation of certain non-local currents in an associated quantised affine algebra \cite{BernardFelder91,IWWZ13}.
The appropriate discretely holomorphic observables have been defined for several types of models, including Ising and 
$Z_N$ models \cite{RajabpourCardy07}, loop models of the Potts \cite{RivaCardy06} and O($N$) types \cite{IkhlefCardy09}, 
the chiral Potts model \cite{IkhlefWeston15}, and more exotic models involving multi-coloured loops \cite{IFC11}.

In a series of papers with C.R.\ Scullard \cite{Jacobsen12,SJ12,Jacobsen13} we have defined a topologically weighted
graph polynomial for Potts and site percolation problems, having properties that are somewhat reminiscent of those found
in the DH approach. This polynomial $P_B$ depends on the Boltzmann weights of the degrees of freedom
living within a ``basis'' $B$, by which we mean a small repeating part of the lattice. The main similarity of the graph polynomial approach
with DH is, that when a set of Boltzmann weights corresponding to an exact solution is inserted, it
produces a root of $P_B$, independently of the size of $B$. In particular, when this size-independence is observed, 
it can be seen as heuristic evidence that we have found an exact solution. An important difference with DH is that
the graph polynomial is defined as a partition function (albeit with some topological weighting of configurations),
unlike the discretely holomorphic observables that take the form of correlation functions.

The graph polynomial method is also practically useful when the model is not exactly solvable. Given a fixed set of physical
coupling constants, let us denote by $P_B(T)$ the evaluation of $P_B$ with Boltzmann weights corresponding to the temperature $T$.
It is then observed \cite{Jacobsen12,SJ12,Jacobsen13,Jacobsen14} that the root $T_B$---that is, a solution of $P_B(T_B) = 0$---converges
very quickly towards the critical temperature $T_{\rm c}$, upon increasing the size of $B$. This extends to models possessing several
critical points \cite{Jacobsen12}, and even to inhomogeneous models with quenched bond disorder (spin glasses) \cite{Ohzeki15}.

The property of $P_B$ just mentioned can then be used as a numerical tool for determining $T_{\rm c}$ very accurately.
This was pursued extensively in \cite{Jacobsen14} for the Potts model defined on all Archimedian lattices, their duals and their medials,
as well as for site percolation on selected lattices (Archimedean and dual Archimedean lattices having only cubic and quartic vertices).

The purpose of this article is to enhance the efficiency of this method, and to place it in a larger perspective by making contact with a
number of existing theoretical constructions. To this end, we consider bases $B$ consisting of $n \times m$ unit cells of the lattice ${\cal L}$
on which the model is defined. According to \cite{SJ12,Jacobsen13}, $P_B$ is defined by endowing $B$ with doubly periodic boundary conditions
and imposing a certain topological weighting of each configuration. The key idea in the present paper is then to take the $m \to \infty$ limit first, so that
the bases effectively become semi-infinite cylinders of circumference $n$. This transforms the criterion $P_B(T_B) = 0$ into an equality
between eigenvalues of two topologically distinct sectors of the corresponding transfer matrix. These eigenvalue problems can then be solved---analytically
for small bases, and numerically for larger ones---within the framework of the periodic Temperley-Lieb algebra \cite{Jacobsen14}.

This construction has several advantages. First, it makes the computation of $T_B = T_{\rm c}(n)$ numerically much more efficient---allowing
basically for doubling the size $n$ attainable, with respect to the previous approach \cite{Jacobsen14}---while maintaining the crucial feature
that $T_{\rm c}(n)$ has no $n$-dependence at all when the model is exactly solvable. Second, it makes useful contact with both the transfer
matrix formalism and with conformal field theory (CFT), allowing for a better understanding of the method. Third, it extends the applicability
of the method beyond Potts and site percolation problems \cite{Jacobsen14}. In particular, we shall show how to adopt it to O($N$) loop models, in both the
dense and dilute phases, in which case the underlying algebra is the  dilute periodic Temperley-Lieb algebra.
Fourth, since twice as many values $T_{\rm c}(n)$ are available for a given problem, the finite-size scaling (FSS) behaviour
can be studied much more carefully, and we can devise extrapolation schemes which are more efficient and reliable than the Bulirsch-Stoer
acceleration of convergence employed in \cite{Jacobsen14}.

We illustrate all these aspects by applying the method to three significant unsolved problems, which in the past have each served as benchmarks
within their respective category:
\begin{enumerate}
 \item Bond percolation on the kagome lattice. This model has been the subject of a long-standing debate, since Wu's ingenious 1979 conjecture
 for the percolation threshold $p_{\rm c}^{\rm Wu} = 0.524\,429\,717\cdots$ \cite{Wu79}. This conjecture was however proved incorrect, both by subsequent
 numerical work \cite{ZiffSuding97}, among which $p_{\rm c} = 0.524\,404\,99 (2)$ \cite{FengDengBlote08} appears to be the most precise value to this date, and---maybe on
 a more fundamental level---by the observation \cite{Jacobsen12} that Wu's conjecture is exactly the outcome of the graph polynomial method
 with the smallest possible $1 \times 1$ basis (containing 6 edges). The previous graph polynomial method \cite{Jacobsen14} gave $p_{\rm c} = 0.524\,404\,999\,173 (3)$,
 a result which we can now improve to
 \begin{equation}
  p_{\rm c} = 0.524\,404\,999\,167\,439 (4) \,.
 \end{equation}
 This exemplifies the first and fourth points made above, revealing in particular that the error bar of \cite{Jacobsen14} was slightly underestimated.
 \item Site percolation on the square lattice. This renowned problem is unsolved essentially because the four-regular square-lattice hypergraph is
 not selfdual. The percolation threshold is in this case given by $p_{\rm c} = 0.592\,746\,05 (3)$ from numerical simulations \cite{FengDengBlote08},
 and by $p_{\rm c} = 0.592\,746\,01 (2)$ from the previous graph polynomial method \cite{Jacobsen14}. We here improve this value to
 \begin{equation}
  p_{\rm c} = 0.592\,746\,050\,792\,10 (2) \,.
 \end{equation}
 \item Self-avoiding polygons (SAP) on the square lattice. This is the polymer ($N \to 0$) limit of a dilute O($N$) loop model on the square lattice,
 in which each vertex can be visited at most once by the polygon. There is a fugacity $z$ per monomer, and no bending rigidity. This model
 has been investigated extensively by exact enumeration techniques \cite{Enting80,ConwayEntingGuttmann93,JensenGuttmann99}.
 The best known critical monomer fugacity is $z_{\rm c} = 0.379\,052\,277\,752 (3)$ \cite{ClisbyJensen12}. We have obtained for this problem
 values of $z_{\rm c}(n)$ up to $n_{\rm max} = 19$. While the extrapolation of these data is compatible with---and more precise than---the
 result of \cite{ClisbyJensen12}, we defer the discussion of extrapolations to a subsequent paper in which several numerical approaches to the
 SAP problem will be compared \cite{GJJS15}.
\end{enumerate}

The paper is organised as follows. In section~\ref{sec:PB} we review the definition of the graph polynomial for the $q$-state Potts model.
The transfer matrix formalism, to be used extensively in this paper, is set up in section~\ref{sec:TM}. Section~\ref{sec:limit} discusses the
$m \to \infty$ limit that leads to the eigenvalue method for the Potts model. Using a few implementational tricks (see section~\ref{sec:practical}),
we can then go on to determine the critical thresholds $p_{\rm c}(n)$ for two selected percolation problems
in section~\ref{sec:res-perco}. The finite-size scaling behaviour of $p_{\rm c}(n)$ is discussed in section~\ref{sec:extrapol}. This leads to
a powerful extrapolation scheme that provides the numerical values of $p_{\rm c} = \lim_{n \to \infty} p_{\rm c}(n)$ given in the abstract.
Examining the relation of the eigenvalue method to conformal field theory, in section~\ref{sec:CFT}, enables us to generalise it to other
representations (see section~\ref{sec:spin-repr}) and other models. In particular, section~\ref{sec:ON} sets up the method for the O($N$) model,
and discusses its relation with exactly solvable models. The application to the SAP problem is also provided there. Finally, section~\ref{sec:disc}
contains a few concluding remarks and perspectives for further investigations.

\section{Graph polynomial}
\label{sec:PB}

We first briefly review the definition of the graph polynomial for the case of the Potts model \cite{Jacobsen12}. To set the scene
for the eigenvalue method, we pay special attention to the ameliorations of computational complexity that were obtained in
\cite{Jacobsen13,Jacobsen14}.

Given a connected graph $G=(V,E)$ with vertex set $V$ and edge
set $E$, the partition function $Z$ of the $q$-state Potts model \cite{Potts52}
can be defined as \cite{FK1972}
\be
 Z = \sum_{A \subseteq E} v^{|A|} q^{k(A)} \,,
 \label{FK_repr}
\ee
where $|A|$ denotes the number of edges in the subset $A$, and $k(A)$
is the number of connected components (including isolated vertices)
in the subgraph $G_A = (V,A)$.
The temperature variable is denoted $v = {\rm e}^K - 1$, where $K$ is the reduced interaction
energy (including the inverse temperature) between adjacent $q$-component spins.
In the representation (\ref{FK_repr}) one can formally allow both $q$ and $v$
to take arbitrary real values.
The special case of bond percolation is obtained
by setting $q=1$ and choosing the probability of an open bond as $p = \frac{v}{1+v}$.

The definition of the graph polynomial $P_B(q,v)$ made initially in \cite{Jacobsen12} was in terms of a
deletion-contraction principle, whose validity it well-known for the partition function itself. The subsequent
work \cite{Jacobsen13} provided an alternative definition that better reveals the topological content of $P_B(q,v)$. We henceforth
suppose that $G$ is an infinite, regular, two-dimensional lattice ${\cal L}$. We define a basis $B$ 
to be a finite subgraph of ${\cal L}$ that produces all of ${\cal L}$ upon application of an appropriate
infinite set of translations that we call the embedding. The definition made in \cite{Jacobsen13} is then in terms of conditioned partition functions,
similar to (\ref{FK_repr}), that are defined on a graph $G=(V,E)$ which is equal to the basis $B$:
\begin{equation}
 P_B(q,v) = Z_{\rm 2D} - q Z_{\rm 0D} \,.
 \label{PB_cluster}
\end{equation}
Here $Z_{\rm 0D}$ is the sum over edge subsets $A \subseteq E$ such that all connected components (clusters)
in the subgraph $G_A = (V,A)$ have trivial homotopy (i.e., are contractible to a point) upon endowing $B$ with
toroidal boundary conditions, whilst $Z_{\rm 2D}$ corresponds to clusters that wrap both periodic directions.
Note that the terms in $Z_{\rm 1D}$, corresponding to clusters wrapping one but not the other periodic direction, do not appear in (\ref{PB_cluster}).

From a computational point of view, the contraction-deletion algorithm
described in \cite{Jacobsen12} has time and memory requirements that grow like $2^{|E|}$, where $|E|$ is the number
of edges in $B$. In practice we shall be interested in bases that are $n \times m$ patterns of the least possible unit cell
for $B$; a multitude of examples was given throughout \cite{Jacobsen14}. 
For a regular lattice ${\cal L}$ one has $|E| = k_{\cal L} n m$, and for Archimedean lattices with the square embedding
considered in \cite{Jacobsen14}, the proportionality constant $k_{\cal L}$ ranges from $4$ (square lattice)
to $9$ (three-twelve and cross lattices); see Table~3 of \cite{Jacobsen14} for details.
In the transfer matrix algorithm of \cite{Jacobsen13} the growth in time and memory is only like $4^{2(n+m)}$, where $2(n+m)$
corresponds to the perimeter (number of terminals) of $B$, that is, the number of vertices shared with translated copies of $B$
within the embedding. Finally, Ref.~\cite{Jacobsen14} provided an improved transfer matrix within the periodic
Temperley-Lieb algebra in which one of the periodic boundary conditions on $B$ was imposed ``on the fly''; this is shown
in Figure~3 of \cite{Jacobsen14}. As a result, the exponential growth was reduced to $4^{2 \, {\rm min}(n,m)}$.

In practical computations, one typically takes $m=n$, and so the computational effort is $2^{k_{\cal L} n^2}$ in \cite{Jacobsen12}, $256^n$ in \cite{Jacobsen13},
and $16^n$ in \cite{Jacobsen14}. Accordingly, the maximum value of $n$ that could be obtained for the Archimedean lattices
was improved from $n_{\rm max} = 2$ in \cite{Jacobsen12} to $n_{\rm max} = 4$ in \cite{Jacobsen13}, and further to
$n_{\rm max} = 7$ in \cite{Jacobsen14}.

In this paper we show how to take the limit $m \to \infty$, so that the basis $B$ becomes a semi-infinite cylinder of circumference
$n$ unit cells of ${\cal L}$. The roots of $P_B(q,v)$ in that limit can then be computed by solving an eigenvalue problem in the periodic
Temperley-Lieb algebra. The corresponding transfer matrix then acts on only $n$ terminals, and time and memory requirements
reduce to $4^n$. Accordingly, the computations can now be taken to $n_{\rm max} = 14$ for the Potts model on the Archimedean lattices.
We shall illustrate this below, by computing the bond percolation threshold on the kagome lattice to high precision. Other values of $q$,
and other lattices, are obviously also of interest, but a more systematic investigation will be reported separately \cite{JS15}.

\section{Transfer matrix}
\label{sec:TM}

We shall refer to (\ref{PB_cluster}) as the Fortuin-Kasteleyn (FK) representation of $P_B(q,v)$.
Each connected component in the subgraph $G_A = (V,A)$ will be called an FK cluster.
As in \cite{Jacobsen14}, we shall need an equivalent formulation in terms of a loop model \cite{BaxterKellandWu76} 
defined on the medial lattice ${\cal M}(B)$. We now review the salient points leading to the definition of the transfer
matrix in the loop representation, as well as the connectivity states that it acts on.

The correspondence between FK clusters and loops can be depicted graphically as follows:
\begin{equation}
\begin{tikzpicture}[scale=0.7]
 \draw[blue,line width=3pt] (1,-1)--(1,1);
 \draw[fill] (1,-1) circle(0.6ex) node[below] {$a$};
 \draw[fill] (1,1) circle(0.6ex) node[above] {$b$};
 \draw[red,line width=1.5pt] (0,1)--(0.8,0.2) arc(45:-45:2mm) -- (0,-1);
 \draw[red,line width=1.5pt] (2,1)--(1.2,0.2) arc(135:225:2mm) -- (2,-1);

\begin{scope}[xshift=5cm]
 \draw[blue,line width=3pt,dashed] (1,-1)--(1,1);
 \draw[fill] (1,-1) circle(0.6ex) node[below] {$a$};
 \draw[fill] (1,1) circle(0.6ex) node[above] {$b$};
 \draw[red,line width=1.5pt] (0,1)--(0.8,0.2) arc(225:315:2mm) -- (2,1);
 \draw[red,line width=1.5pt] (0,-1)--(0.8,-0.2) arc(135:45:2mm) -- (2,-1);
\end{scope}
\end{tikzpicture}
\label{time-like-edge}
\end{equation}
Here $e \equiv (ab) \in E$ is an edge of $G$, and the left (resp.\ right) picture represents the situation where
$e \in A$ (resp.\ $e \notin A$). The equivalent loops (shown in red colour) live on ${\cal M}(B)$;
they bounce off the edge subset $A$ and cut through its complement $E \setminus A$.

To turn this local equivalence into a global one, one uses the Euler relation for a planar graph
to rewrite the partition function (\ref{FK_repr}) in the loop representation as \cite{BaxterKellandWu76}
\be
 Z = q^{|V|/2} \sum_{A \subseteq E} x^{|A|} n_{\rm loop}^{\ell(A)} \,,
 \label{loop_repr}
\ee
where $x = v / \sqrt{q}$, and $\ell(A)$ denotes the number of closed loops induced by the configuration $A$.
The loop fugacity is $n_{\rm loop} = \sqrt{q}$. We shall use the parameters $(q,v)$ and $(n_{\rm loop},x)$ interchangingly.

The loops on ${\cal M}(B)$ provide a representation of the Temperley-Lieb (TL) algebra. Supposing the direction
of ``time'' propagation in (\ref{time-like-edge}) to be upwards, the left (resp.\ right) picture corresponds to the
action of the identity operator ${\sf I}$ (resp.\ the TL generator ${\sf E}_i$) on the two adjacent strands, labelled
$i$ and $i+1$ from left to right. We then have the relations \cite{TemperleyLieb71}
\begin{eqnarray}
 {\sf E}_i^2 = n_{\rm loop} {\sf E}_i \,, \nonumber \\
 {\sf E}_i {\sf E}_{i \pm 1} {\sf E}_i = {\sf E}_i \,, \label{TL} \\
 {\sf E}_i {\sf E}_j = {\sf E}_j {\sf E}_i \quad \mbox{for } |i-j| > 1 \,, \nonumber
\end{eqnarray}
which may be proved graphically by gluing several diagrams in the form (\ref{time-like-edge}) on top of one another.

\begin{figure}
\begin{center}

\begin{tikzpicture}[scale=1.5,>=stealth]
\foreach \xpos in {0,1,2,3}
\foreach \ypos in {0,1,2,3}
 \fill[black!20] (\xpos+0.5,\ypos) -- (\xpos+1,\ypos+0.5) -- (\xpos+0.5,\ypos+1) -- (\xpos,\ypos+0.5) -- cycle;
\foreach \xpos in {0,1,2,3}
\foreach \ypos in {0,1,2,3}
 \draw[black] (\xpos+0.5,\ypos) -- (\xpos+1,\ypos+0.5) -- (\xpos+0.5,\ypos+1) -- (\xpos,\ypos+0.5) -- cycle;

\foreach \xpos in {0,1,2,3}
{
 \draw[fill] (\xpos+0.5,0) circle(0.4ex);
 \draw[fill] (\xpos+0.5,4) circle(0.4ex);
}


\foreach \xpos in {0,1,2,3}
\foreach \ypos in {0,1,2,3}
{
 \draw[red,line width=2pt] (\xpos-0.2,\ypos+0.3) -- (\xpos+0.2,\ypos+0.3);
 \draw[red,line width=2pt] (\xpos+0.8,\ypos+0.3) -- (\xpos+1.2,\ypos+0.3);
 \draw[red,line width=2pt] (\xpos-0.2,\ypos+0.7) -- (\xpos+0.2,\ypos+0.7);
 \draw[red,line width=2pt] (\xpos+0.8,\ypos+0.7) -- (\xpos+1.2,\ypos+0.7);
}

\foreach \xpos in {0,1,2,3}
\foreach \ypos in {0,1,2,3}
{
 \draw[blue,line width=2pt] (\xpos+0.3,\ypos-0.2) -- (\xpos+0.3,\ypos+0.2);
 \draw[blue,line width=2pt] (\xpos+0.7,\ypos-0.2) -- (\xpos+0.7,\ypos+0.2);
 \draw[blue,line width=2pt] (\xpos+0.3,\ypos+0.8) -- (\xpos+0.3,\ypos+1.2);
 \draw[blue,line width=2pt] (\xpos+0.7,\ypos+0.8) -- (\xpos+0.7,\ypos+1.2);
}

\foreach \xpos in {0,1,2,3}
\foreach \ypos in {0,1,2,3}
{
 \draw[red,line width=2pt] (\xpos+0.2,\ypos+0.3) -- (\xpos+0.25,\ypos+0.35);
 \draw[red,line width=2pt] (\xpos+0.8,\ypos+0.3) -- (\xpos+0.75,\ypos+0.35);
 \draw[red,line width=2pt] (\xpos+0.2,\ypos+0.7) -- (\xpos+0.25,\ypos+0.65);
 \draw[red,line width=2pt] (\xpos+0.8,\ypos+0.7) -- (\xpos+0.75,\ypos+0.65);
 \draw[blue,line width=2pt] (\xpos+0.3,\ypos+0.2) -- (\xpos+0.35,\ypos+0.25);
 \draw[blue,line width=2pt] (\xpos+0.7,\ypos+0.2) -- (\xpos+0.65,\ypos+0.25);
 \draw[blue,line width=2pt] (\xpos+0.3,\ypos+0.8) -- (\xpos+0.35,\ypos+0.75);
 \draw[blue,line width=2pt] (\xpos+0.7,\ypos+0.8) -- (\xpos+0.65,\ypos+0.75);
}

\foreach \ypos in {0,1,3}
  \draw (0.5,\ypos+0.5) node{$\check{\sf R}_0$};
\foreach \ypos in {0,1,3}
  \draw (1.5,\ypos+0.5) node{$\check{\sf R}_2$};
\foreach \ypos in {0,1,3}
  \draw (2.5,\ypos+0.5) node{$\cdots$};
\foreach \ypos in {0,1,3}
  \draw (3.5,\ypos+0.5) node{$\check{\sf R}_{2n-2}$};
\foreach \xpos in {0,1,3}
  \draw (\xpos+0.5,2.5) node{$\vdots$};

\draw[very thick,->] (0,-0.5)--(4,-0.5);
\draw (4,-0.5) node[right] {$x$};
\foreach \xpos in {0,1,2,3}
 \draw[thick] (\xpos+0.5,-0.55)--(\xpos+0.5,-0.45);
\draw (0.5,-0.5) node[below] {$0$};
\draw (1.5,-0.5) node[below] {$1$};
\draw (2.5,-0.5) node[below] {$\cdots$};
\draw (3.5,-0.5) node[below] {$n-1$};

\draw[very thick,->] (-0.5,0)--(-0.5,4);
\draw (-0.5,4) node[above] {$y$};
\foreach \ypos in {0,1,2,3}
 \draw[thick] (-0.55,\ypos+0.5)--(-0.45,\ypos+0.5);
\draw (-0.5,0.5) node[left] {$0$};
\draw (-0.5,1.5) node[left] {$1$};
\draw (-0.5,2.5) node[left] {$\vdots$};
\draw (-0.5,3.5) node[left] {$m-1$};
 
\end{tikzpicture}
 \caption{Basis of size $n \times m$ in the loop representation. Terminals of the basis are
 shown as black circles, and periodic boundary conditions have been imposed horizontally.
 The auxiliary and quantum spaces, shown in red and blue colour respectively, sustain loops
 which are acted upon by an $\check{\sf R}_i$-matrix inside each grey square.}
 \label{fig:square-basis-loop}
\end{center}
\end{figure}
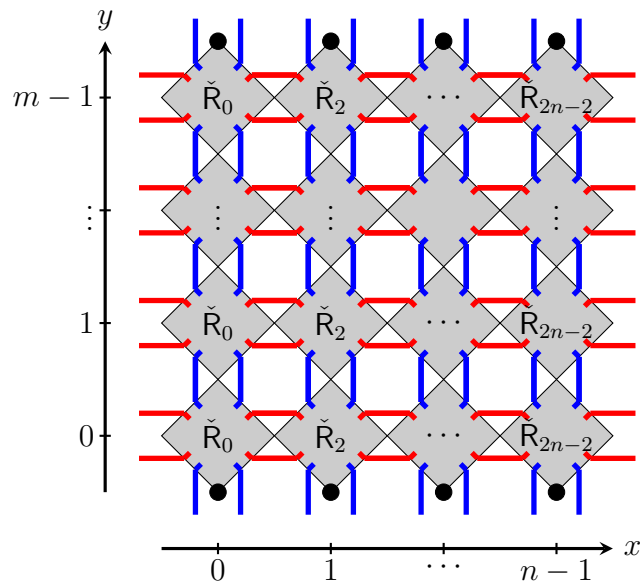

The partition function (\ref{loop_repr}) for a basis of size $n \times m$ can be computed within
the TL algebra as shown in Figure~\ref{fig:square-basis-loop}. The operator $\check{\sf R}_i$ is
here an element of the TL algebra built out of generators ${\sf E}_j$ with $j \in \{i,i+1,i+2\}$, acting
on connectivity states consisting of $2n$ strands, labelled $0,1,\ldots,2n-1$.
The particular arrangement of Figure~\ref{fig:square-basis-loop} is called a four-terminal
representation of $B$ \cite{Jacobsen14}. Further details of this construction and explicit expressions for
$\check{\sf R}_i$ for all Archimedean lattices can be found in \cite{Jacobsen14}.

\begin{figure}
\begin{center}

\begin{tikzpicture}[scale=0.5,>=stealth]

\fill[black!20] (1,4) arc(0:-90:15mm) -- ++(0,1) arc(-90:0:5mm) -- cycle;
\fill[black!20] (6,4) arc(180:270:15mm) -- ++(0,1) arc(270:180:5mm) -- cycle;
\fill[black!20] (2,4) -- (6,0) -- (7,0) arc(180:90:5mm) -- ++(0,1) -- (5,4) -- (4,4) arc(0:-180:5mm) -- cycle;
\fill[black!20] (0,0) -- (1,0) -- (-0.5,1.5) -- ++(0,-1) arc(90:0:5mm) -- cycle;
\fill[black!20] (2,0) arc(180:0:15mm) -- (4,0) arc(0:180:5mm) -- cycle;

\draw (-0.5,0)--(7.5,0);
\draw[dashed] (7.5,0)--(7.5,4);
\draw (7.5,4)--(-0.5,4);
\draw[dashed] (-0.5,4)--(-0.5,0);

\draw[red,line width=1.5pt] (1,4) arc(0:-90:15mm);
\draw[red,line width=1.5pt] (0,4) arc(0:-90:5mm);
\draw[red,line width=1.5pt] (6,4) arc(180:270:15mm);
\draw[red,line width=1.5pt] (7,4) arc(180:270:5mm);
\draw[red,line width=1.5pt] (2,4) -- (6,0);
\draw[red,line width=1.5pt] (7,0) arc(180:90:5mm);
\draw[red,line width=1.5pt] (0,0) arc(0:90:5mm);
\draw[red,line width=1.5pt] (1,0) -- (-0.5,1.5);
\draw[red,line width=1.5pt] (7.5,1.5) -- (5,4);
\draw[red,line width=1.5pt] (4,4) arc(0:-180:5mm);
\draw[red,line width=1.5pt] (2,0) arc(180:0:15mm);
\draw[red,line width=1.5pt] (3,0) arc(180:0:5mm);

\foreach \xpos in {0,1,2,3,4,5,6,7}
{
 \draw[fill] (\xpos,0) circle(0.6ex);
 \draw[fill] (\xpos,4) circle(0.6ex);
}

\draw (0,0) node[below]{$1$};
\draw (1,0) node[below]{$4$};
\draw (2,0) node[below]{$2$};
\draw (3,0) node[below]{$2$};
\draw (4,0) node[below]{$1$};
\draw (5,0) node[below]{$1$};
\draw (6,0) node[below]{$3$};
\draw (7,0) node[below]{$2$};

\draw (0,4) node[above]{$2$};
\draw (1,4) node[above]{$2$};
\draw (2,4) node[above]{$3$};
\draw (3,4) node[above]{$1$};
\draw (4,4) node[above]{$2$};
\draw (5,4) node[above]{$4$};
\draw (6,4) node[above]{$1$};
\draw (7,4) node[above]{$1$};

\draw (3.5,-1) node[below]{(a)};

\begin{scope}[xshift=10cm]

\fill[black!20] (0,0) arc(180:0:5mm) -- cycle;
\fill[black!20] (4,0) arc(180:0:15mm) -- ++(-1,0) arc(0:180:5mm) -- cycle;
\fill[black!20] (3,0) arc(180:90:25mm) -- ++(2,0) -- ++(0,0.5) -- ++(-1.5,0) arc(270:180:10mm) -- ++(-1,0) arc(0:-180:15mm) -- ++(-1,0) -- ++(0,-0.5) arc(0:-90:5mm) -- ++(0,-0.5) arc(90:0:25mm) -- cycle;
\fill[black!20] (2,4) arc(180:360:5mm) -- cycle;
\fill[black!20] (6,4) arc(180:360:5mm) -- cycle;

\draw (-0.5,0)--(7.5,0);
\draw[dashed] (7.5,0)--(7.5,4);
\draw (7.5,4)--(-0.5,4);
\draw[dashed] (-0.5,4)--(-0.5,0);

\draw[red,line width=1.5pt] (0,0) arc(180:0:5mm);
\draw[red,line width=1.5pt] (4,0) arc(180:0:15mm);
\draw[red,line width=1.5pt] (5,0) arc(180:0:5mm);
\draw[red,line width=1.5pt] (3,0) arc(180:90:25mm) -- ++(2,0);
\draw[red,line width=1.5pt] (7.5,3) -- ++(-1.5,0) arc(270:180:10mm);
\draw[red,line width=1.5pt] (4,4) arc(0:-180:15mm);
\draw[red,line width=1.5pt] (0,4) -- ++(0,-0.5) arc(0:-90:5mm);
\draw[red,line width=1.5pt] (-0.5,2.5) arc(90:0:25mm);
\draw[red,line width=1.5pt] (2,4) arc(180:360:5mm);
\draw[red,line width=1.5pt] (6,4) arc(180:360:5mm);

\foreach \xpos in {0,1,2,3,4,5,6,7}
{
 \draw[fill] (\xpos,0) circle(0.6ex);
 \draw[fill] (\xpos,4) circle(0.6ex);
}

\draw (0,0) node[below]{$2$};
\draw (1,0) node[below]{$1$};
\draw (2,0) node[below]{$1$};
\draw (3,0) node[below]{$2$};
\draw (4,0) node[below]{$2$};
\draw (5,0) node[below]{$2$};
\draw (6,0) node[below]{$1$};
\draw (7,0) node[below]{$1$};

\draw (0,4) node[above]{$2$};
\draw (1,4) node[above]{$1$};
\draw (2,4) node[above]{$1$};
\draw (3,4) node[above]{$2$};
\draw (4,4) node[above]{$2$};
\draw (5,4) node[above]{$1$};
\draw (6,4) node[above]{$1$};
\draw (7,4) node[above]{$2$};

\draw (3.5,-1) node[below]{(b)};

\end{scope}

\begin{scope}[xshift=20cm]

\fill[black!20] (0,0) arc(180:0:25mm) -- ++(-1,0) arc(0:180:15mm) -- cycle;
\fill[black!20] (2,0) arc(180:0:5mm) -- cycle;
\fill[black!20] (6,0) arc(180:0:5mm) -- cycle;
\fill[black!20] (0,4) arc(0:-90:5mm) -- ++(0,-1) arc(270:360:15mm) -- cycle;
\fill[black!20] (2,4) arc(180:360:5mm) -- cycle;
\fill[black!20] (4,4) arc(180:270:15mm) -- ++(2,0) -- ++(0,1) arc(270:180:5mm) -- ++(-1,0) arc(0:-180:5mm) -- cycle;

\draw (-0.5,0)--(7.5,0);
\draw[dashed] (7.5,0)--(7.5,4);
\draw (7.5,4)--(-0.5,4);
\draw[dashed] (-0.5,4)--(-0.5,0);

\draw[red,line width=1.5pt] (0,0) arc(180:0:25mm);
\draw[red,line width=1.5pt] (1,0) arc(180:0:15mm);
\draw[red,line width=1.5pt] (2,0) arc(180:0:5mm);
\draw[red,line width=1.5pt] (6,0) arc(180:0:5mm);
\draw[red,line width=1.5pt] (0,4) arc(0:-90:5mm);
\draw[red,line width=1.5pt] (1,4) arc(0:-90:15mm);
\draw[red,line width=1.5pt] (2,4) arc(180:360:5mm);
\draw[red,line width=1.5pt] (5,4) arc(180:360:5mm);
\draw[red,line width=1.5pt] (7,4) arc(180:270:5mm);
\draw[red,line width=1.5pt] (4,4) arc(180:270:15mm) -- ++(2,0);

\foreach \xpos in {0,1,2,3,4,5,6,7}
{
 \draw[fill] (\xpos,0) circle(0.6ex);
 \draw[fill] (\xpos,4) circle(0.6ex);
}

\draw (0,0) node[below]{$2$};
\draw (1,0) node[below]{$2$};
\draw (2,0) node[below]{$2$};
\draw (3,0) node[below]{$1$};
\draw (4,0) node[below]{$1$};
\draw (5,0) node[below]{$1$};
\draw (6,0) node[below]{$2$};
\draw (7,0) node[below]{$1$};

\draw (0,4) node[above]{$2$};
\draw (1,4) node[above]{$2$};
\draw (2,4) node[above]{$1$};
\draw (3,4) node[above]{$2$};
\draw (4,4) node[above]{$1$};
\draw (5,4) node[above]{$1$};
\draw (6,4) node[above]{$2$};
\draw (7,4) node[above]{$1$};

\draw (3.5,-1) node[below]{(c)};

\end{scope}

\end{tikzpicture}
 \caption{Three examples of connectivity states for $n=4$. The numbers along the two time slices provide a canonical coding of the
 connectivity state. Loops are shown as red solid lines. The corresponding FK clusters live in the areas
 shaded in grey, whilst the dual FK clusters live in the white areas.}
 \label{fig:conn-states}
\end{center}
\end{figure}
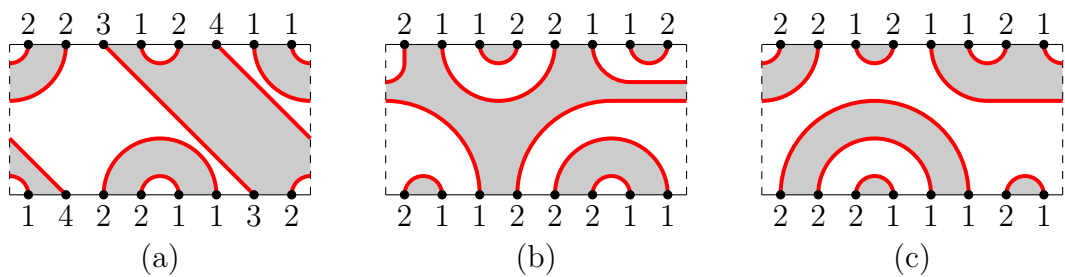

A few examples of connectivity states are shown in Figure~\ref{fig:conn-states}. 
The states are defined on two time slices (top and bottom), that describe the configuration of the system between vertical positions
$y=-1/2$ (bottom) and $y=t-1/2$ (top), after any internal loop has been replaced by the factor $n_{\rm loop}$, upon application of (\ref{TL}).
The transfer matrix that propagates the system from ``time'' $t$ to $t+1$ is then given by the product of all $\check{\sf R}_i$ operators within
a row, where the periodic horizontal boundary conditions are implemented by tracing over the auxiliary spaces (shown in red colour in Figure~\ref{fig:square-basis-loop}).

Let us now be more specific about which variant of the TL algebra we actually need. The choice of periodic boundary conditions in the
horizontal direction means that the standard TL algebra \cite{TemperleyLieb71} (with generators ${\sf E}_i$ for $i=0,1,\ldots,2n-2$) must be made periodic.
In the resulting periodic TL algebra there is an extra generator ${\sf E}_{2n-1}$ that acts across the periodic boundary condition
(i.e., between strands $2n-1$ and $0$). This algebra is however infinite-dimensional, and two modifications (algebra quotients)
must be applied in order to make it finite-dimensional again. First, any loop of non-trivial homotopy (i.e., that winds around the periodic
horizontal direction while being detached from the top and bottom time slices) must be replaced by a factor $n_{\rm wind}$.
Second, loop segments connecting the two time slices are only considered according to which points they connect, and not how many turns
they make around the periodic $x$-direction. The corresponding convention in Figure~\ref{fig:conn-states} is
that among such connecting segments, the one that is leftmost in the top time slice (i.e., that carries the label `3' in Figure~\ref{fig:conn-states})
is required not to cross the periodic direction. Note however that we still need to distinguish whether loop segments that connect a given
time slice to itself crosses the periodic direction or not. With these modifications, the corresponding algebraic object is known
as the {\em augmented Jones-Temperley-Lieb algebra}; see section 3.1 of \cite{GRSV15} for more details.

In order to discuss further the states in Figure~\ref{fig:conn-states}, 
we call {\em arc} a loop segment that connects two points within the
same time slice, and {\em string} a loop segment that connects points on different time slices.
The number of strings $s = 2k$ is always even. When $s > 0$ there are $k$ FK clusters (and also $k$
dual FK clusters) connecting the top and bottom time slices. 
A state can be turned into a pair of ``reduced states'' by cutting the $s$ strings (if any) between the bottom
and top time slices; each reduced state is then associated with only one time slice. Conversely, a pair of reduced
states can be glued along the strings so as to form a ``complete'' state. This gluing can be done in $k$ inequivalent ways,
corresponding to cyclic rotates of the strings of one of the reduced states, in units of two (otherwise the distinction
between FK clusters and dual FK clusters would fail to be respected).
The $s = 0$ reduced states consist only of arcs and can be either {\em closed} (all FK clusters
are bounded away from infinity) or {\em open} (at least one FK cluster is not bounded). 
For example, Figure~\ref{fig:conn-states}b shows a state consisting of two open reduced states, whilst Figure~\ref{fig:conn-states}c depicts
a pair of closed reduced states. Note that each reduced $s=0$ state can be conveniently coded in a binary convention where
the code $1$ (resp.\ $2$) means a arc opening (resp.\ closing).

A subtlety particular to the computation of $P_B(q,v)$ arises because $Z_{\rm 1D}$ does not appear in (\ref{PB_cluster}).
We must therefore set $n_{\rm wind} = 0$. This implies that the $s=0$ states can only be gluings of two open reduced states,
or of two closed reduced states. In other words, mixed gluings are not allowed. We shall call such $s=0$ states open or closed,
respectively.

Below, in section \ref{sec:limit}, we shall expose our main argument for transforming the computation of $P_B(q,v)$ into an eigenvalue
problem in the limit $m \to \infty$. On the level of the transfer matrix, this argument will imply an important simplification
with respect to \cite{Jacobsen14}. Namely, the eigenvalue problem can be solved by using only the top time slice, so that
the transfer matrix acts only on reduced states.
Moreover, only reduced states without strings ($s=0$) are needed---at least in the probabilistic
regime $v > 0$ considered here. There is an equal number of open and closed reduced states, namely
\be
 \frac12 {2n \choose n} \sim 4^n
 \label{num_red_states}
\ee
of each. (In the intermediate stages of the computation, when $n_{\rm aux}$ auxiliary spaces are open, simply replace $n$ by $n + n_{\rm aux}$).
The gain of performance of the present method is largely due to the fact that (\ref{num_red_states}) is much less than the number of
``complete'' states, which grows like $\sim 16^n$; see Eq.~(14) of \cite{Jacobsen14}.

\section{Taking the $m \to \infty$ limit}
\label{sec:limit}

After these preliminaries, we now consider computing $P_B(q,v)$ from (\ref{PB_cluster}) for an $n \times m$ basis with finite $n$ and $m \gg n$;
see Figure~\ref{fig:square-basis-loop}. In this limit, $B$ has the geometry of a semi-infinite cylinder, which naturally suggests an interpretation
in terms of the eigenvalues of a transfer matrix.

When ordering the states according to a decreasing number of strings $s$, the transfer matrix $T$ of section~\ref{sec:TM}, with two
time slices, has a lower block-triangular structure (the blocks being indexed by $s$), since under
the time evolution the number of strings cannot increase. Its eigenvalues are therefore the union of the eigenvalues of each block
on the diagonal, i.e., each eigenvalue can be characterised by the corresponding number of strings $s$. We denote these blocks
by $T^{(s)}$. Moreover, the $s=0$ block is a direct sum of two terms, $T_{\rm open}$ and $T_{\rm closed}$,
corresponding to a pair of open (resp.\ a pair of closed) reduced states. This is so precisely because
contributions to $Z_{\rm 1D}$ are excluded from (\ref{PB_cluster}), implying that loops winding around the cylinder carry the weight
$n_{\rm wind} = 0$. So, as far as the eigenvalue problem is concerned, we can replace $T$ by the direct sum
\be
 \widetilde{T} = \bigoplus_{k =1}^n T^{(s=2k)} \oplus T_{\rm open} \oplus T_{\rm closed} \,.
 \label{decompose_T}
\ee
The modified transfer matrix $\widetilde{T}$ thus has the same spectrum as $T$, and moreover it conserves the quantum number $s$ and, for $s=0$, also the additional quantum number ``open'' and ``closed''. Since $\widetilde{T}$ acts on the top time slice, and unlike $T$ it cannot decrease $s$, it is unable to change the bottom reduced state.
In other words, each of the direct summands $T^{(s)}$ in (\ref{decompose_T})
is in turn a direct sum of $N_s$ of identical blocks, with $N_s$ being the number of reduced states with $s$ strings corresponding to the
bottom time slice. Up to multiplicities of the eigenvalues, it therefore suffices to consider the action of (\ref{decompose_T}) on the reduced
states corresponding to the top time slice. In other words, the set of eigenvalues of $T$ (that acts on ``complete'' states with two time slices) is the union
of eigenvalues of $T^{(s)}$, $T_{\rm open}$ and $T_{\rm closed}$, each restricted to act only on the set of reduced states.%
\footnote{A similar argument has been given in \cite{Richard06,Richard07}.}

We are interested in models with $q >0$, and let us assume further that we are in the ferromagnetic regime, $v > 0$. All Boltzmann weights are
therefore positive, and by the Perron-Frobenius theorem each summand in the decomposition (\ref{decompose_T}) therefore has
a unique, positive largest eigenvalue that we denote $\Lambda^{(s)}$ for $s>0$, respectively $\Lambda_{\rm open}$ and
$\Lambda_{\rm closed}$ for $s=0$. These eigenvalues obviously depend on the size $n$, and on the parameters $(q,v)$.
It follows from the cylinder geometry and the probabilistic assumption $v > 0$ that the eigenvalues are ordered
\be
 \Lambda_{\rm open}, \Lambda_{\rm closed} > \Lambda^{(2)} > \Lambda^{(4)} > \cdots > \Lambda^{(2n)} \,.
 \label{eig_order}
\ee
Moreover, each of the terms in (\ref{PB_cluster}), $Z_{\rm 2D}$ and $Z_{\rm 0D}$, must behave as $\sim \Lambda^m$, where
$\Lambda$ is one of the above eigenvalues. By (\ref{eig_order}) the dominant contributions come from $\Lambda_{\rm open}$
and $\Lambda_{\rm closed}$. Since $Z_{\rm 2D}$ (resp.\ $Z_{\rm 0D}$) corresponds to an FK cluster (resp.\ a dual FK cluster)
spanning the length of the cylinder, we must have
\begin{eqnarray}
 Z_{\rm 2D} &\sim& (\Lambda_{\rm open})^m \,, \\
 Z_{\rm 0D} &\sim& (\Lambda_{\rm closed})^m \,.
\end{eqnarray}
Note that these clusters will also wind around the circumference of the cylinder, with probability $1$ in the limit $m \to \infty$, since the number of strings is $s=0$.

It is obvious that $\Lambda_{\rm open}$ and $\Lambda_{\rm closed}$ are both greater than unity, so $Z_{\rm 2D}$ and $Z_{\rm 0D}$ grow exponentially with $m$.
If (\ref{PB_cluster}) is to have a (positive, unique) zero as a function of $v$---as is indeed observed \cite{Jacobsen12,Jacobsen13,Jacobsen14}---there must
exist some $v_{\rm c}(n) > 0$ so that $\Lambda_{\rm open} = \Lambda_{\rm closed}$.

We can prove this statement as follows.
For $v \gg 1$ the dominant contribution to (\ref{FK_repr}) will be $A = E$, and hence $\Lambda_{\rm open} > \Lambda_{\rm closed}$ by direct computation.
Conversely, for $v \ll 1$ the dominant contribution is $A = \emptyset$, whence $\Lambda_{\rm open} < \Lambda_{\rm closed}$.
Since both terms in (\ref{PB_cluster}) grow exponentially in $m$, the factor of $q$ is unimportant, and the intermediate value theorem implies our main
result
\be
 P_B(q,v) = 0 \quad \Leftrightarrow \quad \Lambda_{\rm open} = \Lambda_{\rm closed} \,,
 \label{main_res}
\ee
valid for a basis $B$ of size $n \times m$, with $n$ finite and $m \to \infty$.

We can therefore find the (unique) positive root, $v > 0$, of the graph polynomial for bases that are semi-infinite cylinders of circumference $n$ by
solving a simple eigenvalue problem over the $s=0$ reduced states. Their number is given by (\ref{num_red_states}) and
grows as $\sim 4^n$, providing a substantial improvement over \cite{Jacobsen14} in which $Z_{\rm 2D}$ and $Z_{\rm 0D}$
were computed by imposing complicated boundary conditions on a transfer matrix acting in the full set of states with
dimension $\sim 16^n$.

In the remainder of this section we first illustrate the main result (\ref{main_res}) in a few simple cases. We end by discussing the
special role of exactly solvable models.

\subsection{Square lattice with $n=1$}
\label{sec:Potts_sq1}

Consider the four-terminal representation of the square lattice with $n=1$. There are $3$ reduced states which
can be written $||$, $()$ and $)($, where $|$ denotes a string, $($ is an arc opening, and $)$ is an arc closing.
For the time being, we allow winding loops with weight $n_{\rm wind}$. It is easy to see that in the basis
$\{ ||, (), )( \}$ the transfer matrix reads $T = T_2 T_1$ with
\begin{eqnarray}
 T_1 &=& \left[ \begin{array}{ccc}
 2x & 0 & 0 \\ x^2 & n_{\rm wind} x^2 & 2 x + n_{\rm loop} x^2 \\ 1 & 2x + n_{\rm loop} & n_{\rm wind} \\
 \end{array} \right] \,, \\
 T_2 &=& \left[ \begin{array}{ccc}
 2x & 0 & 0 \\ 1 & n_{\rm wind} & 2 x + n_{\rm loop} \\ x^2 & 2x + n_{\rm loop} x^2 & n_{\rm wind} x^2 \\
 \end{array} \right] \,,
\end{eqnarray}
where we recall that $n_{\rm loop} = \sqrt{q}$ and $x = v / \sqrt{q}$. Setting $n_{\rm wind} = 0$ we obtain
\be
 T = \left[ \begin{array}{ccc}
 4 x^2 & 0 & 0 \\ n_{\rm loop}+4x & (n_{\rm loop}+2x)^2 & 0 \\ x^4(4+n_{\rm loop} x) & 0 & x^2 (2+n_{\rm loop} x)^2 \\
 \end{array} \right] \,,
\ee
which is indeed a lower block-triangular matrix, corresponding to the blocks $s=2$, $s=0$ (closed), and $s=0$ (open).
Here, each of the blocks have dimension $1$. The (dominant) eigenvalues read
\be
 \Lambda^{(2)} = 4 x^2 \,, \quad
 \Lambda_{\rm closed} = (n_{\rm loop}+2x)^2 \,, \quad
 \Lambda_{\rm open} = x^2 (2+n_{\rm loop} x)^2 \,.
\ee
For $n > 0$ and $x > 0$ the ordering (\ref{eig_order}) is respected indeed.

To investigate the result (\ref{main_res}) we remark that
\be
 \Lambda_{\rm open} - \Lambda_{\rm closed} = n_{\rm loop}(x^2-1)(n_{\rm loop} x^2 + 4 x + n_{\rm loop})
 \label{square_factor}
\ee
is proportional to the graph polynomial $P_B(q,v) = (v^2-q)(q+4v+v^2)$
for the $n \times m = 1 \times 1$ basis \cite{Jacobsen12}.
In this case, where all blocks are one-dimensional, it is obvious---and can be verified by explicit
computations---that the relevant (i.e., positive) roots of $P_B(q,v)$ are independent of $m$.
Note also that, despite of the assumption $v>0$, this example actually correctly describes the
phase diagram of the square-lattice Potts model both in the ferromagnetic ($v>0)$ and antiferromagnetic $(v<0)$ regimes,
i.e., $\Lambda^{(2)}$ can simply be ignored.

\subsection{Kagome lattice with $n=1$}

Consider next the kagome lattice with $n=1$. Setting $n_{\rm wind} = 0$ from the outset and considering
only the $s=0$ states $\{ (),)( \}$ we find that $T$ is a diagonal $2 \times 2$ matrix with eigenvalues
\begin{eqnarray}
 \Lambda_{\rm closed} = (n_{\rm loop}+2x)(n_{\rm loop}^2+4n_{\rm loop} x+6 x^2 + 2 n_{\rm loop} x^3) \,, \\
 \Lambda_{\rm open} = x^2 (2n_{\rm loop} + 12 x + 13 n_{\rm loop} x^2 + 6 n_{\rm loop}^2 x^3 + n_{\rm loop}^3 x^4) \,.
\end{eqnarray}
The difference $\Lambda_{\rm open} - \Lambda_{\rm closed}$ is proportional to
\be
 P_B(q,v) = v^6 + 6 v^5 + 9 v^4 - 2 q v^3 - 12 q v^2 - 6 q^2 v - q^3 \,,
\ee
which is the graph polynomial for the $1 \times 1$ basis. Back in 1979, Wu \cite{Wu79}
conjectured this expression to be the exact critical manifold of the kagome-lattice Potts model,
but our recent work \cite{Jacobsen12,Jacobsen13,Jacobsen14} definitively established that
this is only an approximation corresponding to the smallest possible choice of the basis $B$.

Once again the roots of $P_B(q,v)$ for all the $1 \times m$ bases are independent of $m$,
because the blocks $T_{\rm open}$ and $T_{\rm closed}$ are one-dimensional.

\subsection{Kagome lattice with $n=2$}

In Table~\ref{tab:kagome_vary_m} we show the estimates for the bond percolation threshold $p_{\rm c}$
obtained as the relevant roots of $P_B(1,v)$, with $p = v/(1+v)$, for bases of size $n \times m$ with fixed
$n=2$ and varying $m$. The results for finite $m$ were obtained from the algorithm of \cite{Jacobsen14}.
They are compared with the $m=\infty$ result obtained from (\ref{main_res}) using the transfer matrix
construction of the present paper.


\begin{table}
\begin{center}
 \begin{tabular}{r|l}
 $m$ & $p_{\rm c}$ \\ \hline
  1 & 0.524429717521274793546879681534455071620567416578664793997510 \\
  2 & 0.524406723188231819143234479992589885410333714096742273226669 \\
  4 & 0.524406058417857416583229008103273638077164830301055000364284 \\
  8 & 0.524406057896062955151905518860778390220248322088553687927465 \\
16 & 0.524406057896062634245378836787730760263849423348568828123454 \\
32 & 0.524406057896062634245378836666345666792028877197553627352494 \\
$\infty$ &
        0.524406057896062634245378836666345666792028877197553609980248 \\
 \end{tabular}
 \caption{Bond percolation threshold $p_{\rm c}$ on the kagome lattice using bases of size $2 \times m$ for various $m$.}
 \label{tab:kagome_vary_m}
\end{center}
\end{table}

As expected, the results converge rapidly to the $m=\infty$ limit, the rate of convergence being exponential in $m$. We also note that the
$2 \times \infty$ result $0.524\,406\,057\cdots$ (i.e., the semi-infinite cylinder basis) is closer to the true percolation threshold 
$p_{\rm c} = 0.524\,404\,999\cdots$ \cite{Jacobsen14}
than is the $2 \times 2$ result $0.524\,406\,723\cdots$ (i.e., the square basis). Taking $m \to \infty$ in the $2 \times m$ results is however far from
``gaining one size'', since the $3 \times 3$ result is $0.524\, 405\, 172\cdots$.
These observations extend to arbitrary values of $n$.

\subsection{Exactly solvable cases}

It was shown in \cite{Jacobsen14} that $P_B(q,v)$ factorises over the integers for the three-terminal lattices (square, triangular and hexagonal),
and that in the Ising case $P_B(2,v)$ factorises for any lattice. In these factorisations, one or more ``small'' factors were observed to be independent
of the size $n \times m$ of the basis, and their corresponding roots coincided with known exact solutions. In addition, a few sporadic cases were
found in \cite{Jacobsen14}, mainly concerning $P_B(0,v)$, where a size-independent factorisation occurred, and it was conjectured that these cases would be
exactly solvable.

Because of the $m$-independence of this factorisation result, it should still hold in the $m \to \infty$ limit. We therefore expect that
$\Lambda_{\rm open} - \Lambda_{\rm closed}$ will factorise in exactly solvable cases, spawning an $n$-independent factor whose
roots provide the exact critical points. We have already seen this happen in  (\ref{square_factor}), and we have verified by explicit
computations that this is indeed so also for higher $n$ and for other exactly solvable models.

\section{Practical considerations}
\label{sec:practical}

To take the study to larger sizes $n$, we have implemented numerically the computation of the eigenvalues
$\Lambda_{\rm open}$ and $\Lambda_{\rm closed}$ that enter our main result (\ref{main_res}). The transfer matrix
$T$ needs to be diagonalised in the sectors $T_{\rm open}$ and $T_{\rm closed}$, and the first question to be settled
is which is the most efficient technique for doing so.

We have seen that $T$ can be written as a product of $\check{\sf R}_i$ operators, and these
can in turn be written as product of the elementary operators
\begin{equation}
 {\sf H}_i = {\sf I} + x {\sf E}_i \,, \qquad
 {\sf V}_i = x {\sf I} + {\sf E}_i \,,
 \label{HiVi}
\end{equation}
that add respectively a horizontal (or ``space-like'') and a vertical (or ``time-like'') edge to the lattice \cite{Jacobsen14}.
The operators ${\sf H}_i$ an ${\sf V}_i$ are very sparse, with at most two non-zero entries per column, so the computation
of $w_2 = T w_1$, where $w_1$ is a vector of dimension (\ref{num_red_states}), can be done with time and memory requirements
which are proportional to that dimension. This calls immediately for iterative diagonalisation techniques \cite{Saad}.

The method of choice within this category is the Arnoldi method. However, we shall need to compute the eigenvalues to
high numerical precision (cf.\ Table~\ref{tab:kagome_vary_m}), and we do not know of an implementation of the Arnoldi
method which is compatible with arbitrary precision libraries. Moreover, we need only the largest eigenvalue in each
of the sectors $T_{\rm open}$ and $T_{\rm closed}$, so a very simple method should be sufficient for our purposes.
We have therefore used the most naive scheme, the so-called
power method, in which the operations $w_2 = T w_1$ followed by $w_1 = w_2 / ||w_2||$ are iterated until $||w_2||$ and $w_1$ have converged to the
largest eigenvalue of $T$ and its corresponding eigenvector, respectively.
To obtain convergence of the eigenvalue to 40-digit numerical precision, it turned out necessary to perform several hundreds
of iterations, in particular for large sizes $n$.

The action of Temperley-Lieb generators on the reduced states was described in \cite{Jacobsen14}. Since we have $s=0$ there
are significant simplifications. The insertion and removal of the $n_{\rm aux} = 2$ auxiliary spaces
(made necessary by the four-terminal representation of Figure~\ref{fig:square-basis-loop})
are handled exactly as in \cite{Jacobsen14}. In our implementation the states are stored in a hash table, since we want to take
full advantage of the fact that for some problems (like, for instance, site percolation on the square lattice) the number of states needed can be
even less than (\ref{num_red_states}).

Another major simplification of the eigenvalue method is that no complicated topological considerations---such as those made
in section 3.7 of \cite{Jacobsen14}---will be required in order to distinguish contributions to $Z_{\rm 2D}$ and $Z_{\rm 0D}$ in (\ref{PB_cluster}). 
To choose the sector, it suffices to start the iterative scheme with an initial vector $v$ equal to one of the reduced states in the `open' or
`closed' sector, respectively.

Considerations about efficiency are not limited to choosing the optimal numerical scheme for computing the eigenvalues. Once we can evaluate
the function $f(v) = \Lambda_{\rm open} - \Lambda_{\rm closed}$ for some value of the temperature variable $v$, we need
also an efficient means of adjusting $v$ to its critical value $v_0 \equiv v_{\rm c}(n)$ satisfying $f(v_0) = 0$. Bracketing methods
for finding zeros of a continuous function are numerically very stable, but rather slow. If we allow ourselves to compute derivatives,
we can use instead the Newton-Raphson method, and more generally with $k$'th order derivatives we can employ the $k$'th order
Householder method. The higher-order methods will in principle converge faster, but in practice the computation of high-order
derivatives is numerically unstable, so some compromise must be found.

In practice we have found that the best result is provided by the second-order Householder method, with derivatives being computed by 
the symmetric difference method. Suppose we perform the computations in $d$-digit arithmetics (in practice we have taken $d=40$).
Let us set $\varepsilon = 10^{-d/2}$. Given some estimate $v$ close to $v_0$, one Householder iteration proceeds as follows.
Make three evaluations of $f(v)$,
\begin{equation}
 g_0 = f(v-\varepsilon) \,, \qquad g_1 = f(v) \,, \qquad g_2 = f(v+\varepsilon) \,,
\end{equation}
and apply the following formulae for the finite-difference derivatives:
\begin{equation}
 f_0 = g_1 \,, \qquad
 f_1 = \frac{g_2-g_0}{2 \varepsilon} \,, \qquad
 f_2 = \left( \frac{g_2-g_1}{\varepsilon} - \frac{g_1-g_0}{\varepsilon} \right) \varepsilon^{-1} \,.
\end{equation}
We stress here that to achieve numerical stability, the order of operations has to be carefully respected when computing $f_2$.
Finally, the next approximation to $v_0$ is given by the second-order Householder formula
\begin{equation}
 v_{\rm new} = v - \frac{f_0 f_1}{(f_1)^2 - \frac12 f_0 f_2} \,.
\end{equation}
Extensive tests of this method shows that it has the following nice properties:
\begin{enumerate}
 \item After a few iterations, $v$ converges to $v_0$ to full $d$-digit precision.
 \item If $v$ is chosen sufficiently close to $v_0$, the number of correct digits will double in each iteration.
\end{enumerate}

The main computational effort obviously goes into the largest sizes $n$, and it is important in those cases to provide the
best possible starting value $v_{\rm init}$ for $v$. Thanks to the scaling theory developed in section \ref{sec:extrapol}, we have been
able to predict $v_{\rm init}$ so that $|v_{\rm init} - v_0| < 10^{-15}$ or better. This means that we can attain our $d=40$ digit goal in just
two Householder iterations. In a few cases we have contented ourselves with just a single Householder iteration and a final precision
of at least 30 digits.

\section{Percolation thresholds on selected lattices}
\label{sec:res-perco}

In this section we apply the eigenvalue method to two prominent sample problems which have been extensively studied in the past:
site percolation on the square lattice, and bond percolation on the kagome lattice.

We stress that it would be straightforward to study also the Potts model for other values of $q \neq 1$, or to
switch to any of the other lattices treated in \cite{Jacobsen14}. To this end, it suffices to change
the $\check{\sf R}_i$ matrix to any of the explicit expressions given in \cite{Jacobsen14}, which is a matter
of changing just a few lines of code. We shall however defer these extensions to a future study \cite{JS15}, and
for the time being take the computations for the two problems mentioned as far as possible.

\subsection{Site percolation on the square lattice}
\label{sec:site_square}

The $\check{\sf R}$-matrix for site percolation on the square lattice is given by Eq.~(76) in \cite{Jacobsen14}:
\begin{equation}
 \check{\sf R}_i = {\sf E}_{i+2} {\sf E}_i + v {\sf E}_{i+1} \,,
\end{equation}
where we recall that $q=1$ and the probability of an open bond is $p = \frac{v}{1+v}$.
This $\check{\sf R}$-matrix contains only two out of fourteen possible terms, so thanks to the use of hashing
techniques only a subset of the reduced states will be used in the diagonalisation procedure
(see sections 3.6 and 7.0 of \cite{Jacobsen14} for more details).

The site thresholds were found on $n \times n$ square bases up to $n_{\rm max} = 11$ in \cite{Jacobsen14}.
Using the eigenvalue method we have obtained the thresholds on $n \times \infty$ bases up to $n_{\rm max} = 21$.
These results are shown in Table~\ref{tab:sitesquare}. A comparison with Table 52 of \cite{Jacobsen14} shows that
the $10 \times \infty$ result is very close to the old $11 \times 11$ results, so having taken the $m \to \infty$ limit
can be said, roughly speaking, to have ``gained us one size'' in this case.

\begin{table}
\begin{center}
 \begin{tabular}{r|l}
 $n$ & $p_{\rm c}(n)$ \\ \hline
 1  & 0.5000000000000000000000000000000000000000 \\
 2  & 0.5651977173836393964375280132470308160984 \\
 3  & 0.5888806999178529980514426957517049337221 \\ 
 4  & 0.5914171708531384817988341017359231779642 \\
 5  & 0.5922358232050266776468513240523872931777 \\
 6  & 0.5925073562056416791039647019136652231541 \\
 7  & 0.5926196333998949001725078647635478154618 \\
 8  & 0.5926727605746273159803396143033787878155 \\
 9  & 0.5927006240698093405431688044620515383831 \\
10 & 0.5927163956307984449936472582379354676976 \\
11 & 0.5927258706594202658220083530779420439346 \\
12 & 0.5927318424282431711880689641407018936896 \\
13 & 0.5927357579756109329667635225705210166147 \\
14 & 0.5927384119896431219655807545828528768826 \\
15 & 0.5927402625774169696977289423157479414628 \\
16 & 0.5927415848750128489249007208681102997016 \\
17 & 0.5927425500481430481174634605214209833281 \\
18 & 0.5927432678876343617903095425293033143864 \\
19 & 0.5927438107312915517933469085441350226515 \\
20 & 0.5927442273849199618209333613184304919328 \\
21 & 0.592744551481371482002735520463 \\
\hline
 Ref.~\cite{FengDengBlote08} & 0.59274605 (3) \\
 Ref.~\cite{Jacobsen14} & 0.59274601 (2) \\
 \end{tabular}
 \caption{Site percolation thresholds $p_{\rm c}(n)$ on the square lattice, as computed from $n \times \infty$ bases,
 and two previous results for $p_{\rm c}$.}
 \label{tab:sitesquare}
\end{center}
\end{table}

It is obvious from Table~\ref{tab:sitesquare} that there is agreement with the existing results for $p_{\rm c}$ on at least
the first five digits. More digits can however be obtained by extrapolating the data,
and this will be discussed in section~\ref{sec:extrapol}.

\subsection{Kagome lattice $(3,6,3,6)$}
\label{sec:kagome}

The $\check{\sf R}$-matrix for the Potts model on the kagome lattice is given by Eq.~(29) of \cite{Jacobsen14}:
\begin{equation}
 \check{\sf R}_i = {\sf H}_{i+1} {\sf V}_{i+2} {\sf V}_{i} {\sf E}_{i+1} {\sf V}_{i+2} {\sf V}_i {\sf H}_{i+1} \,,
 \label{eq:Rkagome}
\end{equation}
where the elementary operators ${\sf H}_i$ and ${\sf V}_i$ were defined in (\ref{HiVi}).
We set $q=1$ to obtain the corresponding bond percolation problem.

The bond thresholds were found on $n \times n$ square bases up to $n_{\rm max} = 7$ in \cite{Jacobsen14}.
From the eigenvalue method we have computed the thresholds on $n \times \infty$ bases up to $n_{\rm max} = 14$.
Those results are shown in Table~\ref{tab:kagome}. The value $p_{\rm c}(2)$ was already presented
in Table~\ref{tab:kagome_vary_m}.

\begin{table}
\begin{center}
 \begin{tabular}{r|l}
 $n$ & $p_{\rm c}(n)$ \\ \hline
 1  & 0.5244297175212747935468796815344550716205 \\
 2  & 0.5244060578960626342453788366663456667920 \\
 3  & 0.5244050922187183914064917102789956045159 \\
 4  & 0.5244050138823434506779249332748912013263 \\
 5  & 0.5244050026660985339974686380437997371046 \\
 6  & 0.5244050002521386411660652383853120094089 \\
 7  & 0.5244049995708026048576486896416033403853 \\
 8  & 0.5244049993387487061840419066777093506317 \\
 9  & 0.5244049992479802098067018389586796534330 \\
10 & 0.5244049992084754512628552771194320896813 \\
11 & 0.5244049991897559735118013090108129282307 \\
12 & 0.5244049991802484437799696383468246717858 \\
13 & 0.5244049991751328450538203030184876090453 \\
14 & 0.524404999172242908087780703763 \\
 \hline
 Ref.~\cite{FengDengBlote08} & 0.52440499 (2) \\
 Ref.~\cite{Jacobsen14} & 0.524404999173 (3) \\
 \end{tabular}
 \caption{Bond percolation threshold $p_{\rm c}$ on the kagome lattice, as computed from $n \times \infty$ bases,
 and two previous results for $p_{\rm c}$.}
 \label{tab:kagome}
\end{center}
\end{table}

It is immediately visible that the data in Table~\ref{tab:kagome} converge faster than those in Table~\ref{tab:sitesquare}.
The value of $p_{\rm c}(n)$ with $n=14$ agrees with $p_{\rm c}$ to eleven digits. Looking down the table we can 
also see that the central value for $p_{\rm c}$ given in Ref.~\cite{Jacobsen14} is a bit too high, and that the final extrapolated result is
likely to be slightly below its lower error bound. We shall discuss this extrapolation in section~\ref{sec:extrapol}.

\section{Extrapolations}
\label{sec:extrapol}

We shall now discuss the extrapolation of the data in Tables~\ref{tab:sitesquare}--\ref{tab:kagome} in view of obtaining
final values of the thresholds $p_{\rm c}$ which are as precise as possible.

\subsection{Site percolation on the square lattice}
\label{sec:ss_extrapol}

A reasonable Ansatz for the asymptotic behaviour of $p_{\rm c}(n)$, motivated by general principles of finite-size scaling (FSS),
is that of a series of power-law corrections,
\begin{equation}
 p_{\rm c}(n) = p_{\rm c} + \sum_{k=1}^\infty \frac{A_k}{n^{\Delta_k}} \,,
 \label{FSS0}
\end{equation}
with $0 < \Delta_1 < \Delta_2 < \cdots$. To determine $\Delta_1$ we first form $\delta p_{\rm c}(n) = p_{\rm c}(n) - p_{\rm c}$,
where $p_{\rm c}$ is taken either as the existing best value \cite{FengDengBlote08,Jacobsen14}, or from a preliminary fit to
the data of Table~\ref{tab:sitesquare}. We then consider the sequence
\begin{equation}
 \Delta_1(n) = \frac{\log \big( \delta p_{\rm c}(n) \big) - \log \big( \delta p_{\rm c}(n-1) \big)}{\log(n) - \log(n-1)} \,,
 \label{Delta1ofn}
\end{equation}
which by (\ref{FSS0}) should converge to the first FSS exponent $\Delta_1$.

\begin{figure}
\begin{center}
  \vskip-0.4cm
  \includegraphics[scale=.34]{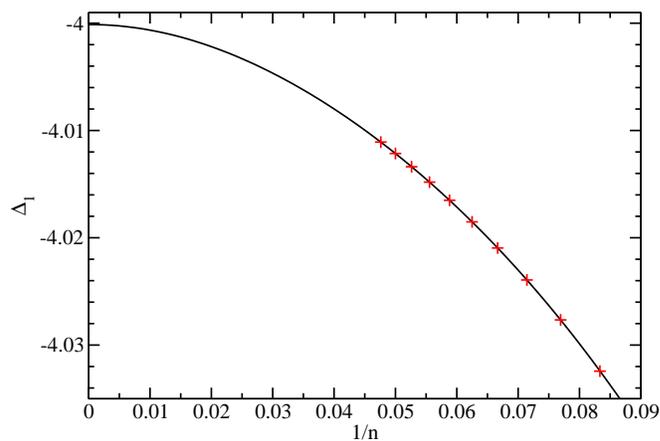}
  \vskip-0.8cm
  \caption{Determination of the first FSS exponent $\Delta_1$ for site percolation on the square lattice.}
  \label{fig:ssDelta1}
\end{center}
\end{figure}

In Figure~\ref{fig:ssDelta1} we show $\Delta_1(n)$ as
a function of $1/n$. The data are very well fitted by a polynomial in $1/n^2$, and allowing for some freedom on the degree of the
polynomial and the number of small-$n$ points to be excluded from the fit, we arrive at the result
\begin{equation}
 \Delta_1 = 4.000\,1 (2) \,.
\end{equation}
This agrees well with the value $w = 4.03 \pm 0.01$ reported in section 7.2 of \cite{Jacobsen14}, which was found as
the optimal choice for the FSS exponent entering the Bulirsch-Stoer algorithm.
It appears inevitable to admit that $\Delta_1 = 4$ exactly.

\begin{figure}
\begin{center}
  \vskip-0.4cm
  \includegraphics[scale=.34]{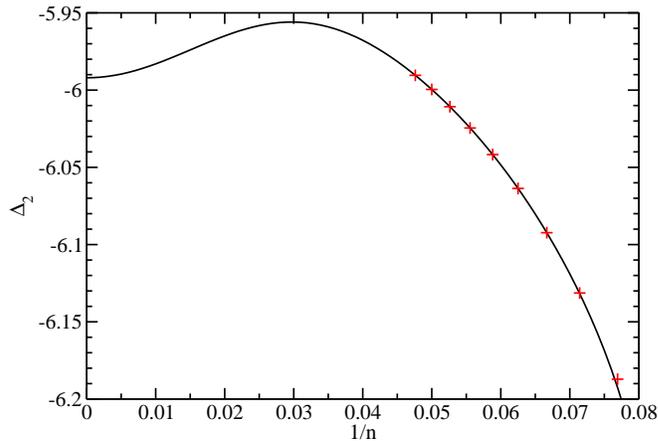}
  \vskip-0.8cm
  \caption{Determination of the second FSS exponent $\Delta_2$ for site percolation on the square lattice.}
  \label{fig:ssDelta2}
\end{center}
\end{figure}

We next form the sequence
\begin{equation}
 p_4(n) = \frac{n^4 p_{\rm c}(n) - (n-1)^4 p_{\rm c}(n-1)}{n^4 - (n-1)^4}
\end{equation}
in which the leading FSS term $A_1/n^4$ has been subtracted off (\ref{FSS0}). From this we form a sequence
\begin{equation}
 \Delta_2(n) = \frac{\log \big( \delta p_4(n) \big) - \log \big( \delta p_4(n-1) \big)}{\log(n) - \log(n-1)} \,,
\end{equation}
which should now converge to the second FSS exponent $\Delta_2$.
We plot $\Delta_2(n)$ against $1/n$ in Figure~\ref{fig:ssDelta2}, along with a polynomial fit in $1/n^2$. This yields
\begin{equation}
 \Delta_2 = 6.00 (1) \,.
 \label{ssDelta2}
\end{equation}
and we henceforth admit that $\Delta_2 = 6$ exactly.

It is now obvious how to continue. In the next round we subtract $A_1/n^4 + A_2/n^6$ from (\ref{FSS0}) and seek to
determine $\Delta_3$ from the residue. Going through the same steps as above we find $\Delta_3 = 8.0 (5)$, and we
conjecture that $\Delta_3 = 8$.

The determinations of $\Delta_1$, $\Delta_2$ and $\Delta_3$ provide compelling evidence that $\Delta_k = 2(k+1)$ for
any $k$. The precise FSS form then reads
\begin{equation}
 p_{\rm c}(n) = p_{\rm c} + \sum_{k=1}^\infty \frac{A_k}{n^{2(k+1)}} \,.
 \label{FSS1}
\end{equation}
We now show how to use this form to obtain a very precise extrapolation for the percolation threshold $p_{\rm c}$.

Let $n_{\rm max}$ denote the largest size for which we have been able to compute $p_{\rm c}(n)$. In the present case
we have $n_{\rm max} = 21$, as seen in Table~\ref{tab:sitesquare}. We first form a series of estimators $p_{M,L}$ in
which the scaling form (\ref{FSS1}) is truncated at the $1/n^M$ term, and in which the data $p_{\rm c}(n)$ is used
up to a maximum size of $n=L$. In other words, we find the unique solution of the linear system
\begin{equation}
 p_{M,L} + \left( \frac{A_1}{n^4} + \frac{A_2}{n^6} + \cdots \frac{A_{M/2-1}}{n^M} \right) = p_{\rm c}(n) \,,
\end{equation}
with $n=L+1-M/2,\ldots,L-1,L$. Second, for a fixed $M$, we form another series of estimators $p_{M}^{(n_0)}$ by
fitting $p_{M,L}$ to the residual dependence predicted by (\ref{FSS1}), but eliminating from the fit the first $n_0$ 
possible values of $L$. That is, we find the unique solution of the linear system
\begin{equation}
 p_{M}^{(n_0)} + \left( \frac{B_1}{n^{M+2}} + \frac{B_2}{n^{M+4}} + \cdots + \frac{B_{n_{\rm max}-n_0-1-M/2}}{n^{2(n_{\rm max}-n_0-1)}} \right) = p_{M,L} \,.
\end{equation}
This is a fit on $n_{\rm max}-n_0-M/2$ different values of $L$ ranging from $1+M/2+n_0$ up to $n_{\rm max}$.

If we eliminate too few data points (i.e., take $n_0$ too small) when forming the estimators $p_M^{(n_0)}$ , the result will be
mediocre because it depends too much on the smallest sizes for which the FSS form (\ref{FSS1}) is dubious.
On the other hand, if we eliminate too many data points (i.e., take $n_0$ too large) the result will again
deteriorate because the fit has too few terms. We would expect an optimum in between these extremes.

\begin{figure}
\begin{center}
  \vskip-0.4cm
  \includegraphics[scale=.34]{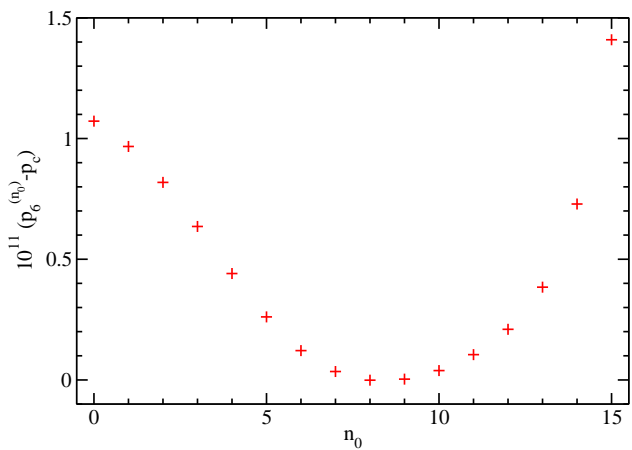}
  \vskip-0.8cm
  \caption{The estimators $p_6^{(n_0)}$ (in arbitrary units) plotted against $n_0$.}
  \label{fig:2ndest}
\end{center}
\end{figure}

In Figure~\ref{fig:2ndest} we show the variation of $p_M^{(n_0)}$ with $n_0$ in the case $M=6$. For reasons of clarity,
we actually plot the quantity $10^{11} (p_M^{(n_0)} - p_{\rm c})$, where $p_{\rm c}$ is our final value for the percolation
threshold, but the units of the ordinate in the plot should really been thought of as arbitrary, since we have not determined $p_{\rm c}$ yet.
We see that there is an extremum (minimum) at some intermediate value of $n_0$, in agreement with the above qualitative
argument. Repeating the plot for other values of $M$ (not shown), it is observed that the minimum becomes more shallow
upon increasing $M$, at least up to a certain point beyond which the quality of the plot deteriorates due to a lack of points.

\begin{table}
\begin{center}
 \begin{tabular}{r|l}
 $M$ & Estimate \\ \hline
 4  & 0.592746050791752 \\
 6  & 0.592746050792111 \\
 8  & 0.592746050792085 \\
10 & 0.592746050792096 \\
12 & 0.592746050792125 \\
14 & 0.592746050792165 \\
16 & 0.592746050792226 \\
 \end{tabular}
 \caption{Estimates $\frac12(p_M^{(8)} + p_M^{(9)})$ for the site percolation threshold on the square lattice $p_{\rm c}$.}
 \label{tab:ss_estimates}
\end{center}
\end{table}

For any value of $M$ in a reasonable range (namely $M=4,6,\ldots,16$) the minimum is found to occur for $n_0 = 8$ or $n_0 = 9$.
The arithmetic mean of those two values of $p_M^{(n_0)}$ can thus be taken as a precise estimate for $p_{\rm c}$. We show these
mean values in Table~\ref{tab:ss_estimates} to 15 significant digits. They are seen to depend only very weakly on $M$ in some
intermediate range (say $M=6,8,10,12$) from which we can extract our final value and error bar for the percolation threshold:
\begin{equation}
  p_{\rm c} = 0.592\,746\,050\,792\,10 (2) \,. \label{ss_final}
\end{equation}

\subsection{Bond percolation on the kagome lattice}

For bond percolation on the kagome lattice we again start by considering an FSS Ansatz of the form (\ref{FSS0}).
It is immediately clear from the data that the leading FSS term does {\em not} correspond to the exponent $\Delta_1 = 4$,
as was found for site percolation on the square lattice. To obtain a unified notation we therefore set $A_1 = 0$ in the kagome
case, and call the leading FSS correction $A_2 / n^{\Delta_2}$.

\begin{figure}
\begin{center}
  \vskip-0.4cm
  \includegraphics[scale=.34]{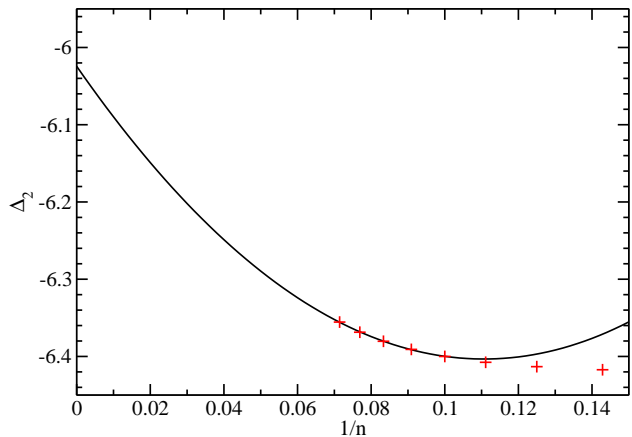}
  \vskip-0.8cm
  \caption{Determination of the leading FSS exponent $\Delta_2$ for bond percolation on the kagome lattice.}
  \label{fig:bkDelta2}
\end{center}
\end{figure}

Figure~\ref{fig:bkDelta2} shows estimates $\Delta_2(n)$ for this leading FSS exponent---extracted from the data $p_{\rm c}(n)$ just as in
(\ref{Delta1ofn})---, plotted against $1/n$. The accompanying fit is a second-order polynomial in $1/n$, and this and similar fits lead to the value
\begin{equation}
 \Delta_2 = 6.00 (5) \,.
\end{equation}
With data up to only $n_{\rm max} = 7$, it was concluded in section 4.3 of \cite{Jacobsen14} that $w \approx 6.36$ was a suitable exponent for the
Bulirsch-Stoer extrapolation. It is clear from Figure~\ref{fig:bkDelta2} how this conclusion could be reached, since effectively only the
the last six data points or so (we have now $n_{\rm max} = 14$) start bending upwards. Although this determination of $\Delta_2$ is somewhat
less accurate than (\ref{ssDelta2}), we are confident in concluding that $\Delta_2 = 6$ exactly.

Comparing the precisions of $\Delta_2$ and $\Delta_3$ obtained in section~\ref{sec:ss_extrapol}, it is clear that there is little hope of
obtaining a convincing determination of $\Delta_3$ in the present case. However, the FSS exponents $\Delta_k$ are not only a property
of these lattice models of percolation, but are also expected to characterise the field theory describing their continuum limit. There is
overwhelming evidence throughout the literature that both models are described, in the continuum limit, by the same conformal field
theory, which can be derived by standard Coulomb gas arguments \cite{JacobsenReview}. So there are good reasons to believe that
we should have $\Delta_k = 2(k+1)$ also in the present case. Different lattice realisations will however give different values of the
non-universal amplitudes $A_k$ in (\ref{FSS0}). It is quite possible that the three-fold rotational symmetry of the kagome lattice
(which replaces the four-fold symmetry of the square lattice) has the effect of setting $A_1 = 0$. This also matches observations made in
\cite{Jacobsen14}.

We therefore proceed with the analysis as in section~\ref{sec:ss_extrapol}, using in particular the scaling form (\ref{FSS1}) with $A_1 = 0$.
Going through the same steps as before we arrive at the final value for the percolation threshold
\begin{equation}
 p_{\rm c} = 0.524\,404\,999\,167\,439 (4) \,. \label{bk_final}
\end{equation}

\section{Connection to conformal field theory}
\label{sec:CFT}

The fact that the leading exponent in the scaling form (\ref{FSS1}) takes a high value, $\Delta_1 = 4$, is responsible for the fast
convergence of $p_{\rm c}(n)$ towards $p_{\rm c}$ and the ensuing precise determinations (\ref{ss_final}) and (\ref{bk_final}).
We now examine how this fast convergence can be linked to considerations about the continuum limit.

The free energy per unit area $f_0(n)$ of a conformally invariant system defined on a semi-infinite cylinder of circumference $n$
scales like \cite{Cardy86,Affleck86}
\begin{equation}
 f_0(n) = f_0(\infty) - \frac{\pi c}{6 n^2} + o \left( n^{-2} \right) \,,
\end{equation}
where $c$ is the central charge of the corresponding conformal field theory (CFT) and $f_0(\infty)$ is the bulk free energy.
This can be related to the largest eigenvalue $\Lambda_0$ of the transfer matrix for a corresponding lattice model
as $f_0(n) = -\frac{\zeta}{n} \log \Lambda_0$, where $\zeta$ is a geometrical factor that depends on the lattice
($\zeta = 1$ for the square lattice) and ensures the correct normalisation per unit area in the lattice model.

Similarly, the free energy $f_i(n)$ of excited states ($i=1,2,\ldots$) has the scaling \cite{Cardy84}
\begin{equation}
 f_i(n) - f_0(n) = \frac{2 \pi x_i}{n^2} + o \left( n^{-2} \right) \,,
 \label{fi_gap}
\end{equation}
where $x_i$ is the corresponding scaling dimension (critical exponent). 
The smallest excitation $f_1(n)$ of the CFT corresponding to percolation is related to the magnetic exponent $x_{\rm m}$,
so we have $x_1 = x_{\rm m}$. The values $c=0$ and $x_{\rm m} = \frac{5}{48}$ are of course known \cite{JacobsenReview},
but they are not important for the following argument.

The important point is that the two transfer matrix sectors, $T_{\rm open}$ resp.\ $T_{\rm closed}$, considered in
section~\ref{sec:limit} correspond to excitations in which an FK cluster (resp.\ a dual FK cluster) is required
to propagate along the semi-infinite cylinder. Define now the corresponding free energies per unit area
\begin{equation}
 f_{\rm open}(n) = -\frac{\zeta}{n} \log \Lambda_{\rm open} \,, \qquad
 f_{\rm closed}(n) = -\frac{\zeta}{n} \log \Lambda_{\rm closed} \,.
\end{equation}
In the continuum limit there is no difference between whether the propagating cluster is an FK cluster or a dual FK
cluster. Therefore $f_{\rm open}(n)$ and $f_{\rm closed}(n)$ both determine the same critical exponent, namely $x_{\rm m}$,
and they both scale like $f_1(n)$ in (\ref{fi_gap}). It follows that the difference
\begin{equation}
 f_{\rm open}(n) - f_{\rm closed}(n) = o \left( n^{-2} \right)
 \label{CFTscaling}
\end{equation}
vanishes fast as $n \to \infty$, right at the critical point $p = p_{\rm c}$. 
 
This is a suggestive argument, but it does not quite explain the convergence properties of the eigenvalue method.
What we have observed in sections~\ref{sec:limit} and \ref{sec:extrapol} is, that if we define a pseudo-critical point
$p_{\rm c}(n)$ as the value of $p$ for which $f_{\rm open}(n) - f_{\rm closed}(n) = 0$, then
\begin{equation}
 p_{\rm c}(n) - p_{\rm c} = O \left( n^{-4} \right) \,,
 \label{Latticescaling}
\end{equation}
and moreover the corrections appear to be $O(n^{-6})$, $O(n^{-8})$, and so on.

It is clear that more work would be required to establish whether (\ref{CFTscaling}) can be shown---obviously using more ingredients---to actually imply (\ref{Latticescaling}).
But one thing that has become clear is, that the eigenvalue method owes its success to the fact that $f_{\rm open}(n)$ and $f_{\rm closed}(n)$
are two different ways of determining the {\em same} critical exponent. This will be exploited further in section~\ref{sec:ON}.

\section{Spin representation}
\label{sec:spin-repr}

It is of interest to review the definition (\ref{PB_cluster}) of the graph polynomial $P_B(q,v)$ when $q \in \mathbb{N}$.
In that case the Potts model can be defined directly in terms of $q$-component spins, instead of the
FK clusters that we have considered this far.

Let again the basis $B$ consist of $n \times m$ unit cells of the lattice ${\cal L}$. Define $Z_{\mu,\nu}$ as the partition function
on $B$ with doubly periodic boundary conditions that are {\em twisted} by $\mu = 0,1,\ldots,q-1$ (resp.\ $\nu = 0,1,\ldots,q-1$) in the
horizontal (resp.\ vertical) direction. By this we mean that the values of a pair of nearest-neighbour spins, $\sigma_i$ and $\sigma_j$, that are
on opposite sides of the horizontal (resp.\ vertical) periodic boundary condition are considered identical if $\sigma_i - \sigma_j = \mu$ mod $q$
(resp.\ $\sigma_i - \sigma_j = \nu$ mod $q$), and different otherwise.

To relate the partition functions $Z_{\mu,\nu}$ in the spin representation to those in the FK-representation ($Z_{\rm 0D}$, $Z_{\rm 1D}$ and $Z_{\rm 2D}$)
we first notice that untwisted boundary conditions are simply doubly periodic, whence
\begin{equation}
 Z_{00} = Z_{\rm 0D} + Z_{\rm 1D} + Z_{\rm 2D} \,.
 \label{Zrel1}
\end{equation}

Consider next the quantity $\sum_{\mu,\nu} Z_{\mu,\nu}$. Configurations in $Z_{\rm 0D}$ contribute to all $q^2$ terms in this sum, whereas those in
$Z_{\rm 2D}$ can only contribute to one term, namely $Z_{00}$. Finally, configurations in $Z_{\rm 1D}$ are such that all clusters that are non-homotopic
to a point have the same topology, i.e., they have the same winding numbers $(n_x,n_y)$ with respect to the horizontal
and vertical periodic boundary conditions \cite{diFrancescoSaleurZuber1987}. These winding numbers are defined up to a global sign change,
$(-n_x,-n_y) \equiv (n_x,n_y)$, and if they are both non-zero they must satisfy
\begin{equation}
 {\rm gcd}(n_x,n_y) = 1 \,.
 \label{gcd1D}
\end{equation}
Now, for a configuration in $Z_{\rm 1D}$ to contribute to $Z_{\mu,\nu}$ we should have
\begin{equation}
 n_x \mu + n_y \nu = 0 \mbox{ mod } q \,.
\end{equation}
Thanks to the constraint (\ref{gcd1D}), this equation has precisely $q$ solutions for the labels $(\mu,\nu)$. Summarising, we have proved that
\begin{equation}
 \sum_{\mu=0}^{q-1} \sum_{\nu=0}^{q-1} Z_{\mu,\nu} = q^2 Z_{\rm 0D} + q Z_{\rm 1D} + Z_{\rm 2D} \,.
 \label{Zrel2}
\end{equation}

Combining (\ref{Zrel1}) and (\ref{Zrel2}) we obtain
\begin{equation}
 Z_{00} - \frac{1}{q} \sum_{\mu=0}^{q-1} \sum_{\nu=0}^{q-1} Z_{\mu,\nu} = \left( 1 - \frac{1}{q} \right) (Z_{\rm 2D} - q Z_{\rm 0D}) \,,
 \label{PB_spin}
\end{equation}
so the quantity on the left-hand side is proportional to the graph polynomial $P_B(q,v)$ by (\ref{PB_cluster}). This was already shown in
the appendix of \cite{Ohzeki15}, by using a more involved argument of duality transformations.

We now consider the $m \to \infty$ limit of (\ref{PB_spin}) in order to obtain an eigenvalue criterion in the spin representation which
is equivalent to (\ref{PB_cluster}). The asymptotic behaviour, as $m \to \infty$, of the partition functions reads
\begin{equation}
 Z_{\mu,\nu} = c_{\mu,\nu} (\Lambda_\mu)^m + c_{\mu,\nu}^{(1)} (\Lambda_\mu^{(1)})^m + \ldots \,,
\end{equation}
where the eigenvalues depend only on $\mu$, but the coefficients can depend on both twist labels.
We have ordered the eigenvalues in decreasing order: $\Lambda_\mu > \Lambda_\mu^{(1)} > \cdots$.
Moreover, the dominant eigenvalues in each sector decrease when the twist increases,
$\Lambda_0 > \Lambda_1 > \ldots$, where obviously $\Lambda_\mu = \Lambda_{q-\mu}$. Moreover,
some of the inequalities might not be sharp when $q$ takes particular values.

Going back to the left-hand side of (\ref{PB_spin}), we see that the dominant contribution $(\Lambda_0)^m$ cancels
out between the two terms. The next-leading contributions come from $\Lambda_0^{(1)}$ and $\Lambda_1$.
These must cancel out in order for $P_B(q,v)$ to vanish:
\be
 P_B(q,v) = 0 \quad \Leftrightarrow \quad \Lambda_0^{(1)} = \Lambda_1 \,,
 \label{main_res_spin}
\ee
valid for finite $n$, in the limit $m \to \infty$. This is the spin-representation version of our main result (\ref{main_res}).

The two eigenvalues involved have a very precise meaning. The leading eigenvalue $\Lambda_0$ in
the untwisted (periodic) sector corresponds to an eigenvector which is invariant under a global permutation
of the spin, $\sigma_i \to p \sigma_i$ with $p \in S_q$. The next-leading eigenvalue $\Lambda_0^{(1)}$ transforms
non-trivially under such a transformation: it picks up a non-trivial $q$'th root of unity. For instance, when $q=2$,
$\Lambda_0^{(1)}$ is the largest eigenvalue corresponding to an eigenvector which is odd under spin reversal.
It is well-known that the free-energy gap between $\Lambda_0^{(1)}$ and $\Lambda_0$ determines the magnetic
exponent $x_{\rm m}$ by (\ref{fi_gap}). On the other hand, $\Lambda_1$ is the largest eigenvalue in the twisted
sector $\mu=1$, in which spin labels are shifted cyclically by one unit when one crosses the periodic boundary condition.
It is equally well-known that its free-energy gap with respect to the ground state $\Lambda_0$ determines 
the {\em same} exponent $x_{\rm m}$. It follows that the criterion (\ref{main_res_spin}) can be discussed in exactly the same
terms as in section~\ref{sec:CFT}.

\section{O($N$) loop model}
\label{sec:ON}

The graph polynomial method, and its eigenvalue version pursued in the present paper, can be seen as a tentative to
generalise the notion of self-duality to situations, where duality is not an exact symmetry. In the Potts model, a duality
transformation exists directly on the lattice, and interchanging the lattice and the dual lattice amounts to shifting cyclically
the sites in the loop representation by one lattice unit. This transformation provides a bijection between the states in
the open and closed sectors, as discussed in section~\ref{sec:TM}.

It is clearly of interest to formulate the graph polynomial method also for other models, and in particular for the O($N$)
loop model \cite{Nienhuis82,BloteNienhuis89}. Although the two models are in the same universality class---to be more precise, the dense phase of the O($N$)
model has the same central charge and a closely related, albeit not identical, operator content as the critical
$q$-state Potts model with $N = \sqrt{q} = n_{\rm loop}$ \cite{JacobsenReview}---their lattice definitions present
subtle differences, which were remarked early on \cite{diFrancescoSaleurZuber1987}. The O($N$) model does not
possess a duality transformation on the lattice, and it treats parity issues in a different way than the loop formulation
of the Potts model. In particular, in the periodic transfer matrix formalism, the O($N$) model can be defined on any number of strands,
whereas in the Potts model the number of strands needs to be even (e.g., there were $2n$ strands in Figure~\ref{fig:square-basis-loop}).
Also, on a semi-infinite cylinder with free boundary conditions at the infinities, the number of non-contractible (winding) loops in
the Potts model must be even, but in the O($N$) model this number can have any parity.

These subtleties have some important consequences in the continuum limit, as can be seen by examining the
operator content of the models in detail. Arguably the most important difference is that the energy operators of
the two models do not coincide, as can be seen from a detailed Coulomb gas (CG) analysis \cite{Nienhuis84,diFrancescoSaleurZuber1987,JacobsenReview}.
The same analysis also reveals that the O($N$) model possesses two kinds of involutions that could be viewed
as duality symmetries. The first involution exchanges the dense and dilute theories corresponding to the same central charge,
by replacing the CG coupling constant $g$ by $1/g$. The second involution originates from the study of modular
invariant partition functions \cite{DSZ87modinv} and amount to exchanging the the role of electric and magnetic
charges in the CG.

\subsection{Eigenvalue method}

\newcommand{\RI}{
\begin{tikzpicture}[scale=0.45]
\draw [black, line width=0.2] (0,-1) -- (1,0); 
\draw [black, line width=0.2] (0,1) -- (1,0); 
\draw [black, line width=0.2] (-1,0) -- (0,-1); 
\draw [black, line width=0.2] (-1,0) -- (0,1);  
\node (1) at (0,-1.5) {$\rho_1$};
\end{tikzpicture}}
\newcommand{\RII}{
\begin{tikzpicture}[scale=0.45]
\draw [black, line width=0.2] (0,-1) -- (1,0); 
\draw [black, line width=0.2] (0,1) -- (1,0); 
\draw [black, line width=0.2] (-1,0) -- (0,-1); 
\draw [black, line width=0.2] (-1,0) -- (0,1);  
\draw[red, line width=0.6mm, rounded corners=7pt] (-0.5,-0.5) -- (-0.1,0.0) -- (-0.5,0.5);
\node (1) at (0,-1.5) {$\rho_2$};
\end{tikzpicture}}
\newcommand{\RIII}{
\begin{tikzpicture}[scale=0.45]
\draw [black, line width=0.2] (0,-1) -- (1,0); 
\draw [black, line width=0.2] (0,1) -- (1,0); 
\draw [black, line width=0.2] (-1,0) -- (0,-1); 
\draw [black, line width=0.2] (-1,0) -- (0,1);  
\draw[red, line width=0.6mm, rounded corners=7pt] (0.5,-0.5) -- (0.1,0.0) -- (0.5,0.5);
\node (1) at (0,-1.5) {$\rho_3$};
\end{tikzpicture}}
\newcommand{\RIV}{
\begin{tikzpicture}[scale=0.45]
\draw [black, line width=0.2] (0,-1) -- (1,0); 
\draw [black, line width=0.2] (0,1) -- (1,0); 
\draw [black, line width=0.2] (-1,0) -- (0,-1); 
\draw [black, line width=0.2] (-1,0) -- (0,1);  
\draw[red, line width=0.6mm, rounded corners=7pt] (-0.5,-0.5) -- (0,-0.1) -- (0.5,-0.5);
\node (1) at (0,-1.5) {$\rho_4$};
\end{tikzpicture}}
\newcommand{\RV}{
\begin{tikzpicture}[scale=0.45]
\draw [black, line width=0.2] (0,-1) -- (1,0); 
\draw [black, line width=0.2] (0,1) -- (1,0); 
\draw [black, line width=0.2] (-1,0) -- (0,-1); 
\draw [black, line width=0.2] (-1,0) -- (0,1);  
\draw[red, line width=0.6mm, rounded corners=7pt] (-0.5,0.5) -- (0.,0.1) -- (0.5,0.5);
\node (1) at (0,-1.5) {$\rho_5$};
\end{tikzpicture}}
\newcommand{\RVI}{
\begin{tikzpicture}[scale=0.45]
\draw [black, line width=0.2] (0,-1) -- (1,0); 
\draw [black, line width=0.2] (0,1) -- (1,0); 
\draw [black, line width=0.2] (-1,0) -- (0,-1); 
\draw [black, line width=0.2] (-1,0) -- (0,1);  
\draw[red, line width=0.6mm, rounded corners=7pt] (-0.5,-0.5) -- (0.5,0.5);
\node (1) at (0,-1.5) {$\rho_6$};
\end{tikzpicture}}
\newcommand{\RVII}{
\begin{tikzpicture}[scale=0.45]
\draw [black, line width=0.2] (0,-1) -- (1,0); 
\draw [black, line width=0.2] (0,1) -- (1,0); 
\draw [black, line width=0.2] (-1,0) -- (0,-1); 
\draw [black, line width=0.2] (-1,0) -- (0,1);  
\draw[red, line width=0.6mm, rounded corners=7pt] (-0.5,0.5)  -- (0.5,-0.5);
\node (1) at (0,-1.5) {$\rho_7$};
\end{tikzpicture}}
\newcommand{\RVIII}{
\begin{tikzpicture}[scale=0.45]
\draw [black, line width=0.2] (0,-1) -- (1,0); 
\draw [black, line width=0.2] (0,1) -- (1,0); 
\draw [black, line width=0.2] (-1,0) -- (0,-1); 
\draw [black, line width=0.2] (-1,0) -- (0,1);  
\draw[red, line width=0.6mm, rounded corners=7pt] (-0.5,-0.5) -- (-0.1,0.0) -- (-0.5,0.5);
\draw[red, line width=0.6mm, rounded corners=7pt] (0.5,-0.5) -- (0.1,0.0) -- (0.5,0.5);
\node (1) at (0,-1.5) {$\rho_8$};
\end{tikzpicture}}
\newcommand{\RIX}{
\begin{tikzpicture}[scale=0.45]
\draw [black, line width=0.2] (0,-1) -- (1,0); 
\draw [black, line width=0.2] (0,1) -- (1,0); 
\draw [black, line width=0.2] (-1,0) -- (0,-1); 
\draw [black, line width=0.2] (-1,0) -- (0,1);  
\draw[red, line width=0.6mm, rounded corners=7pt] (-0.5,-0.5) -- (0,-0.1) -- (0.5,-0.5);
\draw[red, line width=0.6mm, rounded corners=7pt] (-0.5,0.5) -- (0.,0.1) -- (0.5,0.5);
\node (1) at (0,-1.5) {$\rho_9$};
\end{tikzpicture}}

The observations made in section~\ref{sec:CFT} give us an important hint about how to obtain an eigenvalue criterion for the O($N$) loop model,
similar to (\ref{main_res}) and (\ref{main_res_spin}) for the Potts model. It seems that we should try to identify one same critical exponent that arises
in the continuum limit of two topologically distinct sectors of the transfer matrix. Moreover, the introduction to section~\ref{sec:ON} provides
the clue that the two sectors should differ by their charge content (i.e., electric versus magnetic) in the Coulomb gas analysis.

The algebraic framework of the O($N$) model is that of the dilute TL algebra. The precise setup that we shall need is that of the
dilute augmented Jones-Temperley-Lieb algebra, whose description is as in section~\ref{sec:TM}, except that we should now
allow for dilution, in the sense that some vertices and edges are not covered by loops. To make this statement precise, we first
recall that the Potts model defined on a planar graph $G$ is equivalent \cite{BaxterKellandWu76} to a completely packed loop
model defined on the 4-regular medial graph ${\cal M}(G)$, with the loops around a vertex of ${\cal M}(G)$ being in any of the two states (\ref{time-like-edge}).
In the case of the O($N$) model, the loops are defined directly on the chosen graph $G$, which hence needs not be 4-regular \cite{Nienhuis82}, but to keep
things simple we shall concentrate on the O($N$) model defined on the square lattice \cite{BloteNienhuis89}. The two states
(\ref{time-like-edge}) are then replaced by the following nine states of the loops around a vertex
\begin{equation}
\RI \quad \RII \quad \RIII \quad \RIV \quad \RV \quad \RVI \quad \RVII \quad \RVIII \quad \RIX \label{vertices}
\end{equation}
We define $\check{\sf R}_i$ as the sum over those nine diagrams, each one being weighed by the corresponding
Boltzmann weight $\rho_i$ as shown. Integrable choices of $\check{\sf R}_i$ exist \cite{BloteNienhuis89} and will be discussed
in section~\ref{sec:exact_solv_ON},
but for the moment we are interested in the general---and not necessarily critical---case where $\rho_i$ take arbitrary values.

\begin{figure}
\begin{center}

\begin{tikzpicture}[scale=1.5,>=stealth]
\foreach \xpos in {0,1,2,3}
\foreach \ypos in {0,1,2,3}
{
 \fill[black!20] (\xpos+0.15,\ypos+0.15) -- (\xpos+0.85,\ypos+0.15) -- (\xpos+0.85,\ypos+0.85) -- (\xpos+0.15,\ypos+0.85) -- cycle;
}
\foreach \xpos in {0,1,2,3}
\foreach \ypos in {0,1,2,3}
{
 \draw[black] (\xpos+0.15,\ypos+0.15) -- (\xpos+0.85,\ypos+0.15) -- (\xpos+0.85,\ypos+0.85) -- (\xpos+0.15,\ypos+0.85) -- cycle;
}

\foreach \xpos in {0,1,2,3}
\foreach \ypos in {0,1,2,3}
{
 \draw[red,line width=2pt] (\xpos-0.15,\ypos+0.5) -- (\xpos+0.15,\ypos+0.5);
 \draw[red,line width=2pt] (\xpos+0.85,\ypos+0.5) -- (\xpos+1.15,\ypos+0.5);
}

\foreach \xpos in {0,1,2,3}
\foreach \ypos in {0,1,2,3}
{
 \draw[blue,line width=2pt] (\xpos+0.5,\ypos-0.15) -- (\xpos+0.5,\ypos+0.15);
 \draw[blue,line width=2pt] (\xpos+0.5,\ypos+0.85) -- (\xpos+0.5,\ypos+1.15);
}

\foreach \ypos in {0,1,3}
  \draw (0.5,\ypos+0.5) node{$\check{\sf R}_0$};
\foreach \ypos in {0,1,3}
  \draw (1.5,\ypos+0.5) node{$\check{\sf R}_1$};
\foreach \ypos in {0,1,3}
  \draw (2.5,\ypos+0.5) node{$\cdots$};
\foreach \ypos in {0,1,3}
  \draw (3.5,\ypos+0.5) node{$\check{\sf R}_{n-1}$};
\foreach \xpos in {0,1,3}
  \draw (\xpos+0.5,2.5) node{$\vdots$};

\draw[very thick,->] (0,-0.5)--(4,-0.5);
\draw (4,-0.5) node[right] {$x$};
\foreach \xpos in {0,1,2,3}
 \draw[thick] (\xpos+0.5,-0.55)--(\xpos+0.5,-0.45);
\draw (0.5,-0.5) node[below] {$0$};
\draw (1.5,-0.5) node[below] {$1$};
\draw (2.5,-0.5) node[below] {$\cdots$};
\draw (3.5,-0.5) node[below] {$n-1$};

\draw[very thick,->] (-0.5,0)--(-0.5,4);
\draw (-0.5,4) node[above] {$y$};
\foreach \ypos in {0,1,2,3}
 \draw[thick] (-0.55,\ypos+0.5)--(-0.45,\ypos+0.5);
\draw (-0.5,0.5) node[left] {$0$};
\draw (-0.5,1.5) node[left] {$1$};
\draw (-0.5,2.5) node[left] {$\vdots$};
\draw (-0.5,3.5) node[left] {$m-1$};
 
\end{tikzpicture}
 \caption{Transfer matrix construction in the loop representation for the O($N$) model.
 Periodic boundary conditions have been imposed horizontally.
 The auxiliary and quantum spaces, shown in red and blue colour respectively, sustain loops
 which are acted upon by an $\check{\sf R}_i$-matrix inside each grey square.}
 \label{fig:square-basis-ON}
\end{center}
\end{figure}
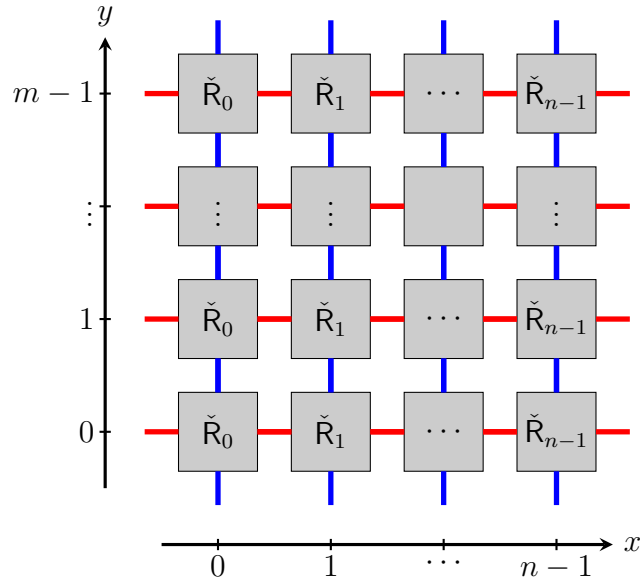

In this dilute model the TL generators ${\sf E}_i$ are defined as before [see (\ref{time-like-edge})], except that ${\sf E}_i$ 
can only be applied when the strands $i$ and $i+1$ are occupied by loop segments, corresponding to the ninth diagram
in (\ref{vertices}). The algebraic relations (\ref{TL}) are unchanged, but must be supplemented by the requirement that
each of the nine generators (\ref{vertices}) can act only when the loop strands have the correct occupancy. This is graphically
clear, but slightly cumbersome to write down in algebraic terms \cite{GrimmPearce93}.
Note that the weight of a closed loop is now denoted $N = n_{\rm loop}$.

The transfer matrix $T$ is again defined as the product over $\check{\sf R}_i$, followed by a trace over the auxiliary space,
as shown in Figure~\ref{fig:square-basis-ON}. As usual in the the algebraic approach to integrable systems, the direction of time propagation is
upwards in diagrams, such as (\ref{vertices}), defining the action of algebra generators, whereas time flows to the North-East when
auxiliary spaces are present. It follows that the diagrams (\ref{vertices}) must be rotated $45^\circ$ in the clockwise direction
before being placed inside the gray squares in Figure~\ref{fig:square-basis-ON}.

The states on which $T$ acts can still be drawn as in Figure~\ref{fig:conn-states}, except that empty vertices are now
possible (and can be represented by the special code `0').

There are two types of operators in the CG. A magnetic operator ${\bf m}_s(r)$ inserts a topological defect, so that
$s$ oriented loop strands are created in a small neighbourhood around the point $r$. The two-point function
$\langle {\bf m}_s (r_1) {\bf m}_{-s} (r_2)\rangle$ corresponds, in the cylinder geometry where $r_1$ and $r_2$ reside
at either extremity, to imposing the propagation of $s$ strings in the transfer matrix setup of section~\ref{sec:TM}.
In particular, the largest eigenvalue $\Lambda^{(s)}$ of $T^{(s)}$ [cf.~(\ref{decompose_T})] determines the critical
exponent $x_s$ of the operator ${\bf m}_s$, via (\ref{fi_gap}).
Note that $s$ can have any parity in the O($N$) model, unlike the Potts case where $s=2k$ must be even.
If the loop weight is parameterised as
\begin{equation}
 N = - 2 \cos(\pi g) \,,
\end{equation}
where $g$ is the CG coupling constant, then \cite{JacobsenReview}
\begin{equation}
 x_s = \frac18 g s^2 - \frac{(1-g)^2}{2g} \,.
 \label{magn_exp}
\end{equation}

The other type of CG operator is the electric operator ${\bf e}_e(r)$, also known as a vertex operator. Its key property is that
the two-point function $\langle {\bf e}_e (r_1) {\bf e}_{-e} (r_2)\rangle$ amounts, in the cylinder geometry and in the
$s=0$ sector, to setting the weight of each non-contractible loop to
\begin{equation}
 N_{\rm wind} = 2 \cos(\pi e) \,.
\end{equation}
The particular choice $e = e_0 \equiv 1 - g$ corresponds to the usual situation $N_{\rm wind} = N$, which provides the ground
state of the model. The corresponding charge $e_0$ is called the background electric charge.
For general $e$, the critical exponent with respect to the ground state reads \cite{JacobsenReview}
\begin{equation}
 \widetilde{x}_e = \frac{e^2 - (1-g)^2}{2g} \,.
 \label{elec_exp}
\end{equation}

The dense (resp.\ dilute) phase of the O($N$) model corresponds to the regime $0 < g \le 1$ (resp.\ $1 \le g \le 2$).
The results for the dense phase also apply to the critical Potts model, by setting $N = \sqrt{q}$.
In particular, setting $N_{\rm wind} = 0$ amounts to forbidding winding loops, so---by the reasoning of sections
\ref{sec:TM} and \ref{sec:CFT}---we obtain the magnetic exponent as $x_{\rm m} = \widetilde{x}_{1/2}$.

Motivated by the introductory remarks in this subsection, we now consider the lowest magnetic excitation ${\bf m}_1$,
corresponding to having one string propagate along the cylinder, with exponent $x_1$ given by (\ref{magn_exp}).
The electric exponent $\widetilde{x}_e$ in (\ref{elec_exp}) can be made to take the same value upon making a particular choice of the
charge $e$:
\begin{equation}
 \widetilde{x}_e - x_1 = \frac{e^2}{2g} - \frac{g}{8} = 0  \quad \Leftrightarrow \quad e = \pm \frac{g}{2} \,.
\end{equation}
This is equivalent to choosing
\begin{equation}
 N_{\rm wind} = \pm \sqrt{2 - N} \,.
 \label{Nwind_ON}
\end{equation}
More generally, we would get $\widetilde{x}_e = x_s$ for $e = \pm s g /2$, but taking the clue from the Potts result, we
should focus on the closest equivalent of the magnetic (order parameter) operator in the Potts model, which is indeed ${\bf m}_1$
in the O($N$) case \cite{Nienhuis84}.

Based on this argument, we define the eigenvalue method for the O($N$) model as follows. For finite size $n$,
find the value of the parameters $\rho_i(n)$ so that the largest eigenvalue in the $s=1$ sector, $\Lambda^{(1)}$, coincides with the
largest eigenvalue $\tilde{\Lambda}$ in the $s=0$ sector with the particular choice (\ref{Nwind_ON}):
\begin{equation}
 \Lambda^{(1)} = \tilde{\Lambda} \,, \quad \mbox{with } N_{\rm wind} = \pm \sqrt{2-N} \,.
 \label{main_res_ON}
\end{equation}
The sign ambiguity on the right-hand side will be resolved later.

This proposed method succeeds or fails depending on whether it can deliver both features that distinguished the eigenvalue method
for the Potts model:
\begin{enumerate}
 \item The values $\rho_i(n)$ should be {\em independent} of $n$ in exactly solvable cases.
 \item For non-solvable cases, $\rho_i(n)$ should converge ``very fast'' in $n$.
\end{enumerate}
This success criterion will be examined in details in the remainder of this section. But let us note for now one
encouraging observation. The dense O($1$) model with vertices (\ref{vertices}) is equivalent to a site percolation 
problem on the square lattice, with certain local interactions depending on $\rho_i$. It is known that for a particular
choice of $\rho_i$, that corresponds to the integrable model \cite{BloteNienhuis89} with an arbitrary inhomogeneous
choice of spectral parameters, the ground state has a combinatorial nature that can be investigated \cite{GarbaliNienhuis14} via the
quantum Knizhnik-Zamolodchikov approach. The ground state in the $s=1$ sector is also combinatorial, and in particular
(\ref{main_res_ON}) holds true for any finite $n$, with the choice $N_{\rm wind} = N = 1$.

\subsection{Exactly solvable cases}
\label{sec:exact_solv_ON}

We now consider the integrable case \cite{BloteNienhuis89,ZhouBatchelor97} of the model (\ref{vertices}) with weights
\begin{eqnarray}
\rho_1(u) &=&1+ \frac{\sin(u) \sin(3\lambda-u)}{\sin(2\lambda) \sin(3\lambda)} \,, \nonumber\\
\rho_2(u) &=& \rho_3(u) = \frac{\sin(3\lambda-u)}{\sin(3\lambda)} \,, \nonumber\\
\rho_4(u) &=& \rho_5(u) = \frac{\sin(u)}{\sin(3\lambda)} \,, \nonumber\\
\rho_6(u) &=& \rho_7(u) = \frac{\sin(u) \sin(3\lambda-u)}{\sin(2\lambda) \sin(3\lambda)} \,, \nonumber\\
\rho_8(u) &=& \frac{\sin(2\lambda-u)\sin(3\lambda-u)}{\sin(2\lambda) \sin(3\lambda)} \,, \nonumber\\
\rho_9(u) &=& -\frac{\sin(u) \sin(\lambda-u)}{\sin(2\lambda) \sin(3\lambda)} \,. \label{ZBweights}
\end{eqnarray}
The spectral parameter $u$ governs the anisotropy of the interactions, and we have here written $\rho_i = \rho_i(u)$ for
later convenience. The crossing parameter $\lambda$ is related to the loop weight via
\begin{equation}
 N = -2 \cos(4 \lambda) \,.
 \label{ZBloop}
\end{equation}
We take arbitrary inhomogeneous spectral parameters, meaning that $u = u_k$ for any vertex in the $k$'th column of the lattice.

For size $n=1$ there is just one reduced state in either of the sectors $s=0$ and $s=1$. The one-dimensional transfer matrices $T^{(s)}$ read
\begin{eqnarray}
 T^{(0)} = \rho_1(u_1) + N_{\rm wind} \rho_6(u_1) \,, \nonumber \\
 T^{(1)} = \rho_7(u_1) + \rho_8(u_1) + \rho_9(u_1) \,.
\end{eqnarray}
Using trigonometic identities, the difference $T^{(0)} - T^{(1)}$ is proportional to $N_{\rm wind} + 2 \cos 2 \lambda$, so we conclude that (\ref{main_res_ON}) is
satisfied with
\begin{equation}
 N_{\rm wind} = -2 \cos(2 \lambda) \,.
 \label{ZBloopwind}
\end{equation}

As $\lambda$ goes from $0$ to $\pi$, the loop weight $N$ runs through the range $[-2,2]$ four times, corresponding to the four branches of
critical behaviour discussed in \cite{BloteNienhuis89}. The first two branches, $\lambda \in [0,\frac{\pi}{4}]$ and $\lambda \in [\frac{\pi}{4},\frac{\pi}{2}]$,
correspond to the dilute and dense phase, respectively.
The relation between (\ref{ZBloop}) and (\ref{ZBloopwind}) is such that the plus (resp.\ minus)
sign in front of the square root in (\ref{main_res_ON}) should be taken for $\lambda \in [\frac{\pi}{4},\frac{3 \pi}{4}]$
(resp.\ $\lambda \in [0,\frac{\pi}{4}] \cup [\frac{3\pi}{4},\pi]$). In particular, the plus (resp.\ minus) sign should be taken for
the dense (resp.\ dilute) phase of the O($N$) model.

We now seek confirmation of these results for size $n=2$. In the sector $T^{(0)}$ there are 3 reduced states which can be written $\circ \circ$, $()$ and $)($,
where $\circ$ denotes an empty site, and the remainder of the notation is as in section \ref{sec:Potts_sq1}. In the sector $T^{(1)}$ the 2 reduced states
are $| \circ$ and $\circ |$. With this ordering of the bases, the transfer matrices read
\begin{equation}
 T^{(0)} = \left[ \begin{array}{ccc}
  \rho_1 \rho_1' + \rho_6 \rho_6' \widetilde{N} \!\!\!\!\!\! &
  \rho_3 \rho_4' N + \rho_4 \rho_3' \widetilde{N} & 
  \rho_3 \rho_4' \widetilde{N} + \rho_4 \rho_3' N \\

  \rho_5 \rho_2' &
  \rho_7 \rho_7' + \rho_9 \rho_8' \widetilde{N} & 
  \rho_8 \rho_8' + \rho_9 \rho_9' + \rho_9 \rho_8' N \!\! \\

  \rho_2 \rho_5' &
  \rho_8 \rho_8' + \rho_9 \rho_9' + \rho_8 \rho_9' N &
  \rho_7 \rho_7' + \rho_8 \rho_9' \widetilde{N} \\
  \end{array} \right]
\end{equation}
and
\begin{equation}
 T^{(1)} = \left[ \begin{array}{ccc}
 \rho_7 \rho_1' + \rho_8 \rho_6' + \rho_9 \rho_6' &
 \rho_2 \rho3' + \rho_5 \rho_4' \\
 \rho_3 \rho_2' + \rho_4 \rho_5' & 
 \rho_1 \rho_7' + \rho_6 \rho_8 + \rho_6 \rho_9' \\
 \end{array} \right] \,,
\end{equation}
where we have abbreviated $\rho_i(u_1) = \rho_i$, $\rho_i(u_2) = \rho_i'$ and $N_{\rm wind} = \widetilde{N}$.
Inserting now (\ref{ZBweights}), (\ref{ZBloop}) and (\ref{ZBloopwind}) we find that the two eigenvalues of $T^{(1)}$
coincide with two of the eigenvalues of $T^{(0)}$, for arbitrary values of the parameters $\lambda$, $u_1$ and $u_2$.

We have similarly studied this model at size $n=3$, in which case ${\rm dim}(T^{(0)}) = 7$ and ${\rm dim}(T^{(1)}) = 6$. Remarkably, we
found that with arbitrary inhomogeneous spectral parameters, $u_1$, $u_2$ and $u_3$, and for arbitrary values of
$\lambda$, all 6 eigenvalues of $T^{(1)}$ were also eigenvalues of $T^{(0)}$. 

The dimensions of these transfer matrices are related to the Motzkin numbers. Define $M(x) = (1+x)(1-3x)$ and consider the generating functions
\begin{eqnarray}
 f_0(x) &=& \frac{1}{\sqrt{M(x)}} = \sum_{n=0}^\infty a_n x^n \,, \nonumber \\
 f_1(x) &=& \frac{2x}{M(x)+(1-x) \sqrt{M(x)}} = \sum_{n=1}^\infty b_n x^n \,.
\end{eqnarray}
Then $a_n = {\rm dim}(T^{(0)})$ and $b_n = {\rm dim}(T^{(1)})$ for a system of size $n$ loop strands.
We have $a_n > b_n$ for $n > 1$. However, both numbers exhibit the same asymptotic behaviour for $n \gg 1$:
\begin{equation}
 a_n \sim b_n \sim \frac12 \left( \frac{3}{\pi n} \right)^{1/2} 3^n \,.
\end{equation}
We conjecture that with inhomogeneous spectral parameters, {\em all} eigenvalues of $T^{(1)}$ are also eigenvalues of $T^{(0)}$,
provided the weight of non-contractible loops is taken as in (\ref{ZBloopwind}).
If true, this would be very promising for finding a genuine graph polynomial for the O($N$) model, i.e., one having properties similar to those of $P_B(q,v)$ in
the Potts case \cite{Jacobsen12,Jacobsen13,Jacobsen14} for {\em finite} $n \times m$ bases, and not just in the $m \to \infty$ limit.
We hope to report more on this soon.

\subsection{Approximation method}

We now investigate the second aspect of the eigenvalue method for the O($N$) model, namely its usefulness as an approximation method
for the critical points of non-solvable models. To this end, we apply it to the problem of self-avoiding polygons (SAP) on the square lattice,
which is the $N \to 0$ limit of a loop model in which each occupied edge has the weight $z$. There is no bending rigidity, and the osculating
vertices $\rho_8$ and $\rho_9$ are disallowed. The Boltzmann weights (\ref{vertices}) are thus
\begin{eqnarray}
 \rho_1 &=& 1 \,, \nonumber \\
 \rho_2 &=& \rho_3 = \rho_4 = \rho_5 = \rho_6 = \rho_7 = z \,, \nonumber \\
 \rho_8 &=& \rho_9 = 0 \,.
 \label{SAP_weights}
\end{eqnarray}
Note that we have suppressed the spectral parameter $u$, since this model is not integrable.

This SAP model has been extensively studied by exact enumeration techniques \cite{Enting80,ConwayEntingGuttmann93,JensenGuttmann99},
and the critical monomer fugacity is known to very high precision \cite{ClisbyJensen12}
\begin{equation}
 z_{\rm c} = 0.379\,052\,277\,752 (3) \,.
\end{equation}
This value corresponds to the smallest $z > 0$ for which the generating function, as obtained by exact enumeration, exhibits a singularity.
In the formulation of the problem in terms of a partition function, with the weights (\ref{SAP_weights}), this corresponds to selecting the
dilute branch of the O($N$) model, and implies taking the minus sign in (\ref{main_res_ON}). We therefore set $N=0$ and $N_{\rm wind} = -\sqrt{2}$.

\begin{table}
\begin{center}
 \begin{tabular}{r|l}
 $n$ & $z_{\rm c}(n)$ \\ \hline
 2  & 0.3832870437289217825415444959209990643484 \\
 3  & 0.3800152822923947541103727449094743052839 \\
 4  & 0.3793419092420152604076859124268482909456 \\ 
 5  & 0.3791615386298805591124869699564102732536 \\
 6  & 0.3791017465104568577033096312174651793134 \\
 7  & 0.3790779263723816763857349117326710080035 \\
 8  & 0.3790669419366251682820022783255996752011 \\
 9  & 0.3790612863965732376129739341339159714858 \\
10 & 0.3790581237478262657302859193323348704028 \\
11 & 0.3790562392439348634338963536547147709970 \\
12 & 0.3790550583590770828697993099179842253186 \\
13 & 0.3790542873705249946097446478792002255473 \\
14 & 0.3790537664746062070854620937409756594548 \\
15 & 0.3790534041836437725305420784870138649786 \\
16 & 0.3790531458388626510867578645132848654379 \\
17 & 0.3790529575762840825464391257666224019613 \\
18 & 0.3790528177462476184521578790271596607432 \\ 
19 & 0.3790527121228867470 \\ 
 \hline
 Ref.~\cite{ClisbyJensen12} & 0.379052277752 (3) \\
 \end{tabular}
 \caption{Critical fugacities $z_{\rm c}(n)$ of the SAP model on the square lattice, as computed from $n \times \infty$ bases,
 and a previous result for $z_{\rm c}$.}
 \label{tab:tonyON}
\end{center}
\end{table}

By diagonalising the transfer matrices $T^{(0)}$ and $T^{(1)}$ and proceeding as in section~\ref{sec:practical}, we have obtained
values of $z_{\rm c}(n)$ up to $n_{\rm max} = 19$ for the SAP problem. They are shown in Table~\ref{tab:tonyON}.
It is clear that these data exhibit the required fast convergence, and a detailed analysis---to be presented elsewhere \cite{GJJS15}---reveals that
the scaling is in fact compatible with (\ref{FSS1}).

We should emphasise that the formulation in terms of a partition function allows us to study the problem (\ref{SAP_weights}) also for other
values of $N$, and in the dense phase. The dilute $N=1$ case corresponds to an Ising model on the square lattice in which configurations
of alternating spins ($+-+-$ or $-+-+$) around a lattice face have been disallowed, since we have set $\rho_8 = \rho_9 = 0$. This produces
a value of the critical domain wall fugacity, $z_{\rm c} = 0.421\,326\cdots$, which is slightly lower than that of the standard Ising model, 
which reads $z_{\rm c}^{\rm Ising} = (1 + \sqrt{2})^{-1} = 0.414\,214\cdots$.

The {\em dense} phase of SAP (with $N=0$) does not appear to have been studied explicitly within the exact enumeration framework, but the corresponding
$z_{\rm c}$ is likely to manifest itself as a subdominant singularity of standard, dilute SAP.

We finally note that when applying (\ref{main_res_ON}) to a non-solvable model, the parameter $z_{\rm c}(n)$ can indeed be tuned so that the
leading eigenvalues of $T^{(0)}$ and $T^{(1)}$ coincide, but the remaining eigenvalues of $T^{(1)}$ will in general not be equal to eigenvalues of
$T^{(0)}$.

\section{Discussion}
\label{sec:disc}

In this paper we have transformed the graph polynomial method of \cite{Jacobsen12,Jacobsen13,Jacobsen14} into an eigenvalue method.
This corresponds formally to taking the $m \to \infty$ limit of the $n \times m$ bases $B$ that enter the definition of the graph polynomial $P_B$.
The advantages of this reformulation are numerous and have been discussed in the Introduction. We can add to this list that, on a technical level,
the eigenvalue method requires only the reduced states in the transfer matrix setup (see section~\ref{sec:TM}) and avoids the rather complicated
topological considerations of \cite{Jacobsen14} (see section 3.7 of that reference in particular), two facts that make the practical implementation
of the method considerably easier.

On a more fundamental level, the eigenvalue formulation has revealed that the method hinges on identifying two distinct topological sectors of the
transfer matrix that lead to the determination of one same critical exponent. This has enabled us to extend the applicability of the method from
the $q$-state Potts model---including bond and site percolation problems---to encompass O($N$) models in various phases, even in the presence of inhomogeneities.
Also in the O($N$) case we have demonstrated that the method is both capable of
\begin{enumerate}
 \item detecting exact solvability (in the sense that integrable models lead to results independent of the size $n$), and of
 \item generating approximations to the critical parameters that converge rapidly in $n$.
\end{enumerate}

This first aspect poses a set of fundamental questions that should motivate future research. In particular, the possible link between exact factorisation
in the graph polynomial method and discrete holomorphicity (alias conservation of non-local currents in quantised affine algebras \cite{BernardFelder91,IWWZ13}),
or related manifestations of exact solvability, remains to be elucidated. With the present extension from Potts to O($N$) models, we have demonstrated
that the graph polynomial method is likely to be as ubiquitous and versatile as discrete holomorphicity. The fact that the O($N$) version of the method
required us to impose a very particular value of $N_{\rm wind}$ in (\ref{main_res_ON}), and the ensuing massive eigenvalue coincidences between $T^{(1)}$ and $T^{(0)}$,
are strongly reminiscent of phenomena encountered in representation theory \cite{LCFT_review}, and this possible link should be examined as well.

The second aspect has enabled us to study the finite-size scaling properties of the method in much more detail than \cite{Jacobsen14}. In particular,
we have developed powerful extrapolation schemes capable of determining the critical point to within 15-digit precision.
In future work, we plan to extend these determinations to other lattices (following \cite{Jacobsen14} in the Potts case), and to
other values of the parameters $q$ and $N$. From a more practical perspective, we are working on a parallel implementation of the algorithm
which should make accessible larger $n$ and lead to even higher precision \cite{JS15}.

Finally, the extension to yet other models, such as $Z_N$ models and multi-coloured loops,
previously considered from the discrete holomorphicity perspective \cite{RajabpourCardy07,IFC11}, should also be investigated.
The question of whether the present method applies to loop models with non-trivial boundary interactions \cite{JacobsenSaleur08,DJS09,DJS10} provides another
appealing perspective.

\section*{Acknowledgments}

This work was supported by the Agence Nationale de la Recherche
(grant ANR-10-BLAN-0414:~DIME) and the Institut Universitaire de
France. The author warmly thanks A.J.\ Guttmann, M.\ Ohzeki and C.R.\ Scullard for discussions and collaboration on related subjects,
and A.D.\ Sokal for the kind permission to use computational resources (provided by Dell Corporation) at New York University.
He is also grateful for the hospitality of the Centre of Excellence for Mathematics and Statistics of Complex Systems (Melbourne University)
and the Galileo Galilei Institute of Theoretical Physics (Arcetri, Florence) where part of this work was accomplished.

\section*{References}
\bibliographystyle{iopart-num}
\bibliography{SJ}

\end{document}